\documentstyle[10pt,epsfig,amssymb,float,twoside]{article}
\newcommand{\module}[1] { {\mid\! #1\! \mid} }
\newcommand{\esp}       { \;\;\;\; }
\newcommand{\refp}[1]   {(\ref{#1})}

\renewcommand{\thebibliography}[1]{\section*{References}\list
{[\arabic{enumi}]}{\settowidth\labelwidth{[#1]}\leftmargin\labelwidth
\advance\leftmargin\labelsep
\usecounter{enumi}}
\def\newblock{\hskip .11em plus .33em minus .07em}
\sloppy\clubpenalty4000\widowpenalty4000
\sfcode`\.=1000\relax}

\begin{document}

\centerline{\large \bf A new construction for spinor wave equations}
\centerline{
Samuel {\sc de Toro Arias}$^{1,2}$\footnote{E-mail~: {\tt sdetoro@spec.saclay.cea.fr}}
and Christian {\sc Vanneste}$^1$\footnote{E-mail~: {\tt vanneste@ondine.unice.fr}. Author for
correspondence} 
}
\begin{center}
$^1$ {\it Laboratoire de Physique de la Mati\`ere Condens\'ee, CNRS UMR $6622$,\\
Universit\'e de Nice-Sophia Antipolis, Parc Valrose, B.P. $71$, F-$06108$ Nice 
Cedex $02$, France.}\\
$^2$ {\it Service de Physique de l'Etat Condens\'e, Centre d'Etudes de Saclay,\\
l'Orme des Merisiers, F-$91191$ Gif sur Yvette Cedex, France.}
\end{center}

\vspace{0.5cm}
\begin{flushleft}
Ref. B7104, accepted for publication in EPJ B \\
Shortened version of the title~: {\bf A new construction for spinor wave equations}
\end{flushleft}

\vspace{0.5cm}
\begin{flushleft}
PACS~: $03.65$.Pm -- Relativistic wave equations \\
PACS~: $02.70$.-c -- Computational techniques \\
\end{flushleft}

%**************************************************************************

\begin{abstract}
The construction of discrete scalar wave propagation equations in arbitrary 
inhomogeneous media was recently achieved by using elementary dynamical 
processes realizing a discrete counterpart of the Huygens principle. In this 
paper, we generalize this approach to spinor wave propagation. Although the 
construction can be formulated on a discrete lattice of any dimension, for 
simplicity we focus on spinors living in $1+1$ space-time dimensions. The 
Dirac equation in the Majorana-Weyl representation is directly recovered by 
incorporating appropriate symmetries of the elementary processes. 
The Dirac equation in the standard representation is also obtained by using its 
relationship with the Majorana-Weyl representation. \\

\centerline{{\bf R\'esum\'e}}  
La construction d'\'equations discr\`etes pour des ondes scalaires se propageant
dans un milieu arbitrairement h\'et\'erog\`ene a \'et\'e realis\'ee r\'ecemment 
par l'introduction de processus dynamiques \'el\'ementaires qui ob\'eissent \`a un \'equivalent
discret du principe de Huygens. Dans cet article, nous g\'en\'eralisons cette approche
\`a la propagation d'ondes spinorielles. Bien que la construction puisse \^etre formul\'ee
sur un r\'eseau discret de dimension quelconque, nous nous limitons au cas simple
de spineurs d\'efinis sur un espace-temps de dimension $1+1$. L'\'equation de 
Dirac en repr\'esentation de Majorana-Weyl est retrouv\'ee en imposant aux processus
\'el\'ementaires certaines sym\'etries appropri\'ees. L'\'equation de Dirac
en repr\'esentation standard est \'egalement obtenue \`a partir de sa
relation avec la repr\'esentation de Majorana-Weyl.
\end{abstract}

%**************************************************************************

\section{Introduction}
The formulation of wave propagation by using the Huygens principle
was investigated some years ago by the Transmission Line Matrix Modeling method 
(TLM)~\cite{johns}. This method retrieves the Maxwell equations by introducing current 
and voltage impulses which propagate along the bonds and are scattered on the nodes of a 
mesh of transmission lines. Refinements lead the TLM method to describe complex 
boundary conditions, gain or losses and propagation in inhomogeneous media for 
electromagnetic waves~\cite{hoefer}. Such an idea was renewed recently for studying 
time-dependent wave propagation for scalar waves in inhomogeneous 
media~\cite{vanneste,sebbah}. Instead of considering currents or voltage pulses, the authors 
introduced some arbitrary scalar quantities that propagate on a Cartesian lattice. Both approaches 
include the action-by-proximity of the pulses, or of the scalar quantities, 
when they propagate from node to node in one time step and the emission of secondary 
waves at each node by means of a scattering process which emits the incident energy in all 
directions. Since voltage and current impulses are equivalent to electric and magnetic fields 
on a two-dimensional mesh, it is not surprising that the TLM method applies to the Maxwell 
equations. Nevertheless, the nature of the wave equations which result from such a formulation 
applied to scalar quantities propagating on a Cartesian lattice 
remains unclear. A beginning of answer arises in the work of Sornette {\it et al.}~\cite{sornette}, 
who have enlarged the scalar model to construct the Klein-Gordon equation and the 
Schr\"odinger equation. However, the results were limited to homogeneous media and the 
Schr\"odinger equation was ill-defined in the continuum limit. In order to generalize this attempt, 
a constructive approach was explored recently~\cite{detoro} to recover various kinds of scalar 
waves in an inhomogeneous medium. Starting from basic principles and incorporating fundamental 
symmetries to the scattering nodes, the authors derived in a systematic way time-dependent scalar 
wave equations in inhomogeneous media. They exhibited a unified equation which properly tuned 
by a unique parameter yields either the Klein-Gordon equation or the Schr\"odinger equation with 
a well defined continuum limit. This derivation offers a general framework, including the 
related TLM approach and opens up possible generalizations to describe spinor or vector 
wave propagation. 

In the present paper, we extend this previous work to the description
of discrete wave propagation equations for free spinor fields. For simplicity, we restrict the
presentation to the derivation of the Dirac equation on a one-dimensional regular lattice 
but the proof can be worked out on any underlying discrete lattice. Throughout the paper, we
call currents the scalar quantities which obey a simple dynamics of propagation and scattering.  
The problem being linear, we define naturally the two scalar components of the spinor field as 
linear superpositions of those currents. In order to derive 
coupled propagation equations linking together both components of the spinor field, we first require 
closure conditions compelling the form of the matrices which describe the scattering processes.
Finally, by taking into account appropriate space-time relativistic symmetries, the discretized 
Dirac equation in $1+1$ dimensions is recovered. The current model is constructed for the two 
usual representations of the Dirac equation in $1+1$ space-time dimension, i.e. the 
Majorana-Weyl representation and the standard representation.

Independently from the above approach, different kinds of microscopic models 
describing spinor wave equations have been proposed in the literature. Those models belong
to two distinct classes~: they are either based on a random walk description, 
leading to Euclidean-invariant spinor wave equations, 
or they rely on a complex hopping-type dynamics on a lattice, leading to 
relativistic-invariant spinor wave equations. The first class of microscopic 
probabilistic models deals with critical behavior of the free Majorana spinor 
field where the time variable $t$ is turned into a 
parameter that controls the approach to a critical point (the system is 
at criticality when $t \rightarrow \infty$). It is indeed well 
known~\cite{itzykson,sokal} that local scalar field theories can be 
studied within a random walk representation, where
the time $t$ corresponds to the length of a brownian path. Among such 
approaches, Mc Keon {\it et al.}~\cite{ord} 
have constructed a probabilistic model, based on binomial processes, to 
describe the spinor field in $1+1$ space-time dimensions. As the Dirac 
equation is recovered from real stochastic processes, their 
method applies only to the real Majorana representation of the Dirac equation. 
Another random walk approach, including a spin factor, is developed 
in~\cite{itoi}. The model is defined on three-dimensional 
lattices and exhibits different critical behavior depending on the value 
of the spin. The second class of models deals with the Dirac equation itself. 
The paradigm of those lattice models are the Susskind fermions~\cite{susskind}
which have been extensively used in lattice gauge 
theories. A different but closed approach is exposed in~\cite{zee90}, 
where the model describes the dynamics of point-like spinless particles 
on a three-dimensional square lattice in the presence of complex hopping 
rates of modulus unity. Related works have been applied to study condensed 
matter topics such as the ground state of the Heisenberg antiferromagnet in two 
or three dimensions~\cite{zee91}, or the energy spectrum of flux 
states~\cite{kunszt,pryor}. The construction in terms of currents presented in 
this paper can be considered as belonging to that second class of models.

The backbone of the paper are sections~\ref{maj} and~\ref{sta}, where we
describe the construction of the Dirac equation in the 
Majorana-Weyl and standard representations respectively. Sections~\ref{dyna_maj} and
\ref{dyna_sta} introduce the basis of the current model in both representations. 
Sections~\ref{eq_maj}, \ref{eq_sta}, deal with the derivation of the discrete coupled 
propagation equations. The Majorana-Weyl construction turns out to be solvable by choosing 
suitable symmetries in section~\ref{symet_maj}. The current model is achieved by computing 
the scattering matrices in section~\ref{scatt_maj}. On the contrary, the
standard representation cannot be constructed in the same direct way. Hence, to compute the
scattering matrices we use the linear transformation linking both representations 
(section~\ref{symet_sta}). Finally, section~\ref{concl} is devoted to discuss further 
issues and applications of this work.

%*****************************************************************************

\newpage
\section{Majorana-Weyl representation}~\label{maj}
\subsection{Basic definitions~: currents and fields}~\label{dyna_maj}
This section is devoted to the construction of the Dirac equation
in the Majorana-Weyl representation, which reads (see~\cite{zuber})

\begin{equation}
i\frac{\partial \Psi}{\partial t} = \left[-ic \sigma_3 
\frac{\partial}{\partial x} + \frac{m c^2}{\hbar} \sigma_1 \right] \Psi,
\label{eq:contdirac}
\end{equation}
where $\Psi=(\psi_L \:\: \psi_R)^{T}$ is the two-component spinor field 
and $\sigma_i, i=1,2,3,$ denotes the Pauli matrices.

\begin{figure}[hbt]
\centerline{\epsffile{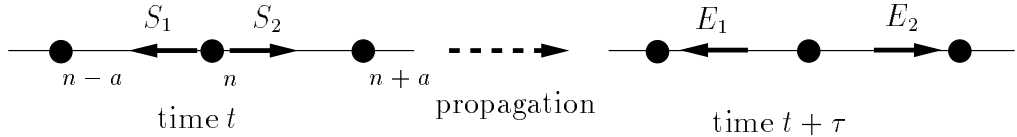}}
\caption{Sketch of the propagation step. The outgoing currents $S_{i}$, $i=1,2,$
on node $n$ propagate in one time step to become incident currents $E_{j}$, $j=1,2,$
on the neighbor nodes.}
\label{fig:prop_scalar}
\end{figure}

The elementary bricks of the construction are called currents. The currents 
are complex numbers which propagate from node to node along the bonds of a 
Cartesian lattice, each bond carrying two currents propagating in opposite 
directions. For simplicity, the lattice is chosen to be one-dimensional and 
regular with a lattice mesh size $a$. At any time $t$, the system is completely 
defined by the values of all currents along the chain. The time variable is 
also discrete and $\tau$ denotes the time unit. A current propagates 
between two neighboring nodes in one time step. All the currents propagate 
simultaneously, i.e. the outgoing currents, denoted $S_{i}$, $i=1,2$, leave the 
nodes at some time $t$ and become incident currents on the neighboring nodes 
at time $t+\tau$. The incident currents are denoted $E_{i}$, $i=1,2$ 
(Fig.\ \ref{fig:prop_scalar}). 

\begin{figure}[hbt]
\centerline{\epsffile{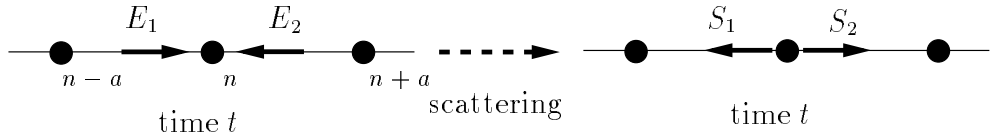}}
\caption{Sketch of the scattering step. The incident currents $E_{i}$, $i=1,2,$ on node $n$
are scattered instantaneously to become outgoing currents $S_{j}$, j=1,2.}
\label{fig:scatt_scalar}
\end{figure}

After this propagation step, all the incident currents are 
instantaneously scattered. The scattering process is described by scattering 
matrices attached to each node. It transforms the incident currents $E_{i}$, 
$i=1,2$, on one node into outgoing currents $S_{j}$, $j=1,2$ on the same node 
(Fig.\ \ref{fig:scatt_scalar}).

\begin{figure}[hbt]
\centerline{\epsffile{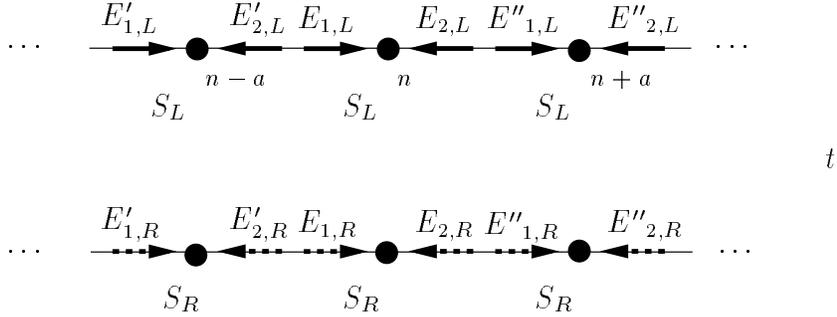}}
\caption{The chain is split into two subchains. The upper and lower subchains 
carry the $L$ and $R$ currents respectively.}
\label{fig:modele_maj1}
\end{figure}

\begin{figure}[hbt]
\centerline{\epsffile{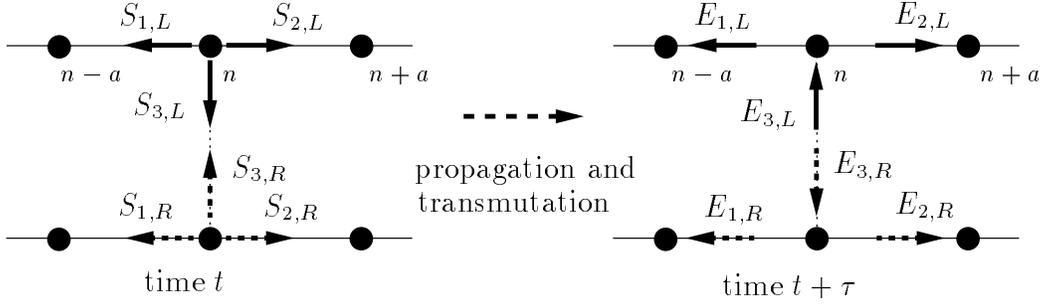}}
\caption{Propagation step for the $L$ and $R$ currents.}
\label{fig:modele_maj2}
\end{figure}

Since our aim is to describe the two chiral fields 
$\psi_L$ and $\psi_R$, two kinds of currents, $L$ and $R$ currents, are 
naturally introduced, leading to four currents propagating simultaneously on 
each bond. For clarity, the one-dimensional lattice can be pictured 
by two sublattices carrying the $L$ and $R$ currents respectively 
(Fig.\ \ref{fig:modele_maj1}).
\newpage
Due to the structure of the Dirac equation~\refp{eq:contdirac}
which displays a coupling term between the two chiral 
components $\psi_L$ and $\psi_R$, we also introduce additional 
currents which propagate between the two sublattices in one time step as 
depicted in Figure~\ref{fig:modele_maj2}. 

\begin{figure}[hbt]
\centerline{\epsffile{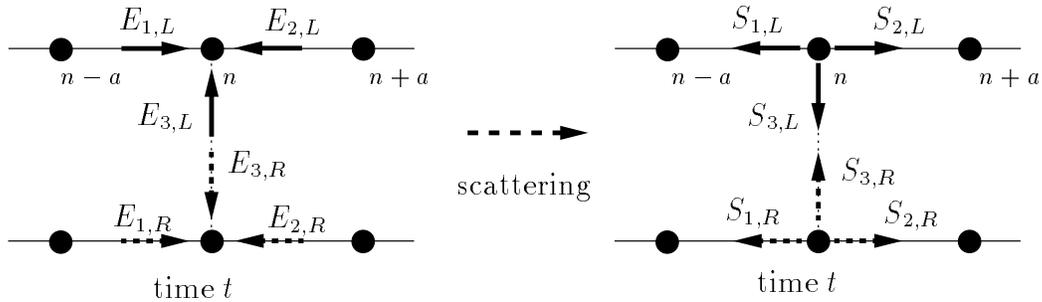}}
\caption{Scattering step for the $L$ and $R$ currents.}
\label{fig:modele_maj3}
\end{figure}

These additional currents, which are called the 
$L$ or $R$ node-current in the following, do not propagate 
along the chain but remain located on the same node. They can 
be considered as propagating along a internal time axis which couples 
the $L$ and $R$ sublattices at each node of the chain. 
Moreover, these node currents participate in the same scattering process as 
the propagating currents (Fig.\ \ref{fig:modele_maj3}).

Thus, the scattering process reads

\begin{eqnarray}
S_{i,L}(n,t) & = & \sum_{j=1}^{3} s_{ij,L} E_{j,L}(n,t), 
\esp i=1,2,3,\label{eq:scatt_majL}\\
S_{k,R}(n,t) & = & \sum_{l=1}^{3} s_{kl,R} E_{l,R}(n,t),
\esp k=1,2,3.\label{eq:scatt_majR}
\end{eqnarray}
where $s_{ij,L}$ and $s_{kl,R}$ are the complex entries of the  scattering 
matrices $S_L$ and $S_R$  attached to each node. The ensemble of all the 
scattering matrices belonging to the chain defines the medium or the 
``background'' in which the currents live.
Once the scattering matrices are known, the evolution of the system at any time is completely
determined given any initial values of the currents. Throughout the paper the 
scattering matrices are chosen to be identical from node to node indicating 
that the medium is uniform. However, the matrices $S_L$ and $S_R$ do not need to be 
identical.

The dynamics includes all the features of a discrete counterpart of the 
well-known Huygens principle. Namely, the principle of action-by-proximity 
from one node to the nearest-neighboring nodes is taken into account 
by the propagation step, while the interferences and  
the radiation of the energy in all directions are taken into account 
by the scattering process.

Let us introduce some convenient notations. The bonds attached to a given 
node are numbered as displayed in Figure~\ref{fig:modele_maj3} : $1$ for the left 
bond, $2$ for the right bond and $3$ for the node bond. 

\begin{figure}[hbt]
\centerline{\epsffile{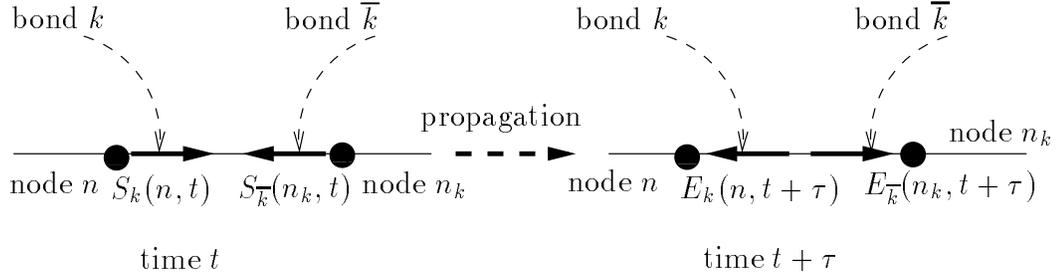}}
\caption{Notations for the bond linking the central node $n$ to one of its neighbor 
node $n_{k}$ and the propagating currents defined on this bond.}
\label{fig:def_bond}
\end{figure}
\begin{figure}[hbt]
\centerline{\epsffile{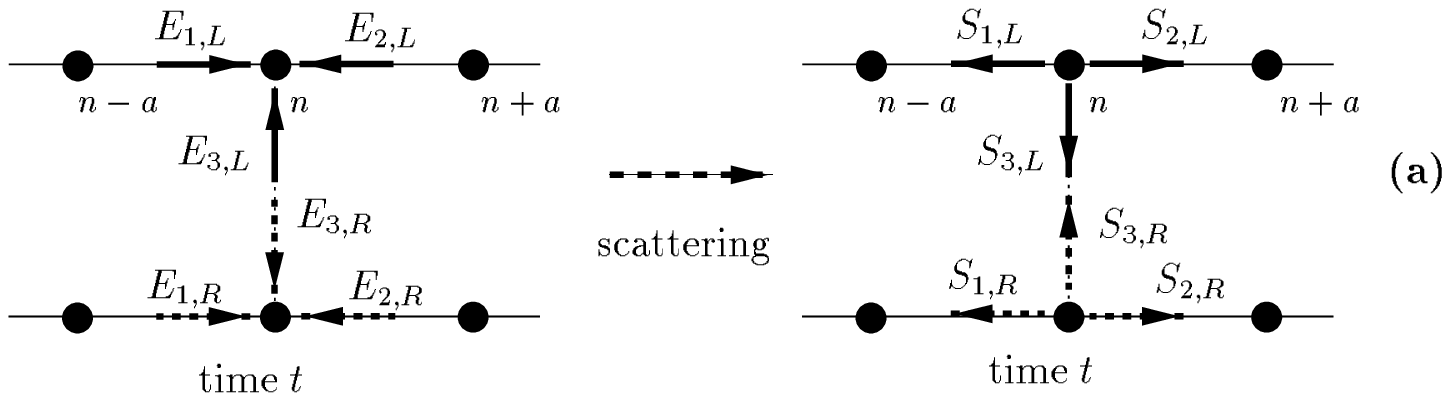}}
\centerline{\epsffile{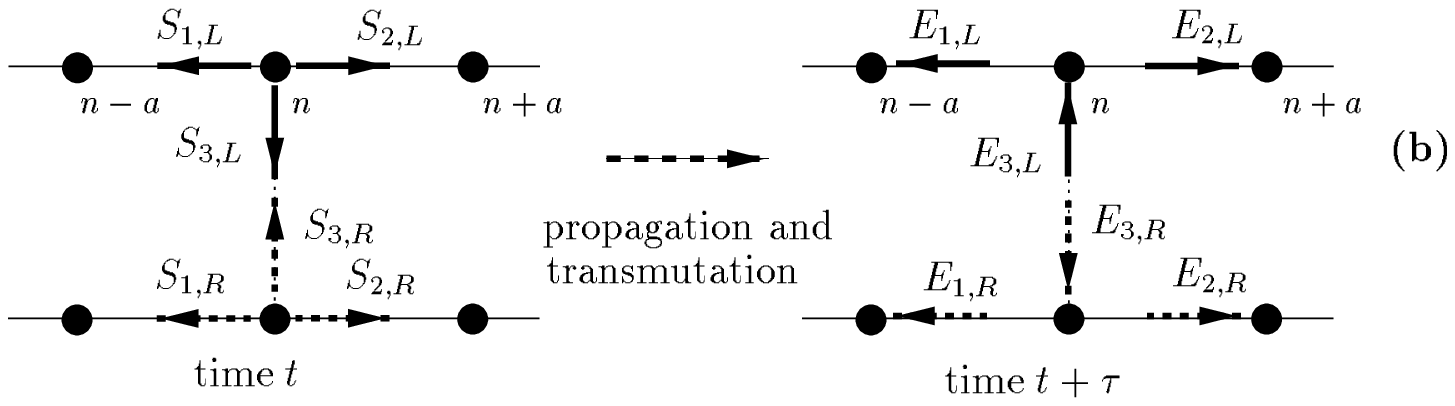}}
\caption{Scattering step {\bf (a)} and propagation step {\bf (b)} for the 
$L$ and $R$ currents.}
\label{fig:difprop_chiral}
\end{figure}

Now consider a current $S_k(n,t)$ leaving the node $n$ along the bond $k$, 
$k=1,2,3$ (Fig. \ref{fig:def_bond}). At the next time step, this current becomes an input 
current $E$ on the neighbor node along the $k^{\mbox{$\scriptstyle{th}$}}$
direction. This neighbor node is labelled ${n_k}$. 
Note that this notation implies ${n_3}=n$. Moreover, viewed from node ${n_k}$, 
the bond on which lies the input current $E$
is denoted $\overline{k}$, leading to labelling this current as $E=E_{\overline{k}}(n_{k},t+\tau)$. 
This notation implies $\overline{1}=2$, $\overline{2}=1$, $\overline{3}=3$. Hence, 
the relations linking the outgoing and the incident currents during the propagation step read

\begin{eqnarray}
E_{k,L}(n,t+\tau) & = & S_{\overline{k},L}(n_k,t), \esp k=1,2, 
\label{eq:proprule_chiral1} \\
E_{k,R}(n,t+\tau) & = & S_{\overline{k},R}(n_k,t), \esp k=1,2,
\nonumber \\
E_{3,L}(n,t+\tau) & = & S_{3,R}(n,t),
\label{eq:proprule_chiral2} \\
E_{3,R}(n,t+\tau) & = & S_{3,L}(n,t). \nonumber
\end{eqnarray}
The propagation rules \refp{eq:proprule_chiral2} take into account a 
transmutation of the node currents $S_{3,L}$ and $S_{3,R}$ into $E_{3,R}$
and $E_{3,L}$ respectively. Figure~\ref{fig:difprop_chiral} summarizes the two steps
defining the dynamics of the $L,R$ currents.

To pursue the construction of the Dirac equation within this two-species
current model, the chiral fields need to be defined. 
The fields $\psi_L$, $\psi_R$ are complex valued functions which 
are defined on the nodes of the chain. The linearity of the 
problem compels the choice for each chiral field $\psi_L$ or $\psi_R$ 
to be a linear superposition of the $L$ or $R$ currents respectively.
As the outgoing currents can be expressed as a function of the incident currents
according to the scattering equations~\refp{eq:scatt_majL}, \refp{eq:scatt_majR}
it is sufficient to define the chiral fields in terms of the incident currents solely~:

\begin{eqnarray}
\psi_L(n,t) & = & \sum_{k=1}^{3} \lambda_{k,L} \, E_{k,L}(n,t), 
\label{def:chiral_comp_a}\\
\psi_R(n,t) & = & \sum_{l=1}^{3} \lambda_{l,R} \, E_{l,R}(n,t),
\label{def:chiral_comp_b}
\end{eqnarray}
where $\lambda_{k,L}$, $\lambda_{l,R}$ are complex numbers.

\subsection{Discrete propagation equations 
linking $\psi_L$ and $\psi_R$}~\label{eq_maj}
Under certain assumptions, which we call thereafter closure conditions, 
we show that the two chiral fields, $\psi_L$, $\psi_R$, satisfy the 
following discrete propagation equations

\begin{eqnarray}
\psi_L(n,t+\tau) & = & f_L\left( \psi_L(n',t'),\psi_R(n',t'),
\psi_L(n'',t'',\psi_R(n'',t''),\ldots \right),
\label{eq:wave_field_L} \\
\psi_R(n,t+\tau) & = & f_R\left( \psi_R(n',t'),\psi_L(n',t'),
\psi_R(n'',t''),\psi_L(n'',t''),\ldots \right),
\label{eq:wave_field_R}
\end{eqnarray}
where the fields, $\psi_L$, $\psi_R$, on node $n$ and at time $t+\tau$ are 
functions of both chiral fields on the same node and/or on nodes 
$n'$, $n''$, $\ldots$, at previous times $t'$, $t''$, $\ldots$  
Equations \refp{eq:wave_field_L}, \refp{eq:wave_field_R} are closed equations
in the fields. This means that $f_{L,R}$ are functions involving the two 
chiral fields only and do not depend explicitly on the incident currents 
of both kinds. Moreover, the number of terms in the right hand side of 
equations~\refp{eq:wave_field_L}, 
\refp{eq:wave_field_R} must be finite. Those two closure conditions 
strongly constrain the form of the scattering 
matrices $S_{L,R}$, so that one finds

\begin{eqnarray}
S_{L} & = & P_{L} +\mbox{diag}(\mu_{1,L},\mu_{2,L},\mu_{3,L}),
\label{def:chiral_scatt_a} \\
S_{R} & = & P_{R} +\mbox{diag}(\mu_{1,R},\mu_{2,R},\mu_{3,R}),
\label{def:chiral_scatt_b}
\end{eqnarray}
where

\begin{equation} 
(P)_{ij,\alpha} = \rho_{i,\alpha}\lambda_{j,\alpha}, \esp i,j=1,2,3, \esp \alpha=R,L,
\label{def:chiral_scatt_c}
\end{equation} 
and $\lambda_{j,\alpha}$ denotes the complex numbers appearing in the definitions of the 
chiral fields. The $\rho_{i,\alpha}$ and $\mu_{i,\alpha}$ are complex numbers to be determined. 
As usual $\mbox{diag}$ denotes a diagonal matrix. The discussion leading to the 
form~\refp{def:chiral_scatt_a}, 
\refp{def:chiral_scatt_b} of the scattering matrices is exactly the same as for scalar 
fields~\cite{detoro}. Therefore,
we have only quoted the result here. According to equations~\refp{def:chiral_scatt_a} 
and~\refp{def:chiral_scatt_b}, it is easy to check that the outgoing current $S_k$ for 
each species splits into a term proportional to the incident current $E_{k}$ of the same 
kind on the same bond $k$ and into a term proportional to the field of the same kind

\begin{eqnarray}
S_{k,L} & = & \rho_{k,L} \psi_{L} + \mu_{k,L} E_{k,L},  \label{eq:outcurrent} \\
S_{k,R} & = & \rho_{k,R} \psi_{R} + \mu_{k,R} E_{k,R}, \esp k=1,2,3. \label{eq:outcurrentR} 
\end{eqnarray}
As the constructions of the equations~\refp{eq:wave_field_L}, \refp{eq:wave_field_R} 
obeyed by $\psi_{L}$ and $\psi_{R}$ are carried out in the same way, we only 
consider the equation for $\psi_L$ in the following. An analogous equation can
be derived for the field $\psi_R$. We fix the space-time units, $a$ and $\tau$, to unity. 
Let us consider the $L$ field on node $n$ at time $t+1$ which is defined as 
(Eq.~\refp{def:chiral_comp_a})

\[
\psi_L(n,t+1) = \sum_{k=1}^{3} \lambda_{k,L} \, E_{k,L}(n,t+1).
\] 
According to the propagation step, it can be expressed in terms of the outgoing currents 
at the previous time $t$

\begin{equation}
\psi_L(n,t+1) = \sum_{k=1,2} \lambda_{k,L} \, 
S_{\overline{k},L}(n_k,t) + \lambda_{3,L} \, S_{3,R}(n,t). 
\label{eq:psiL_int_1}
\end{equation}
In the remainder of this derivation, we will use 
drawings as often as possible since those representations are far more 
transparent than formal equations. For example, 
the definition of the $L$ field (Eq.~\refp{def:chiral_comp_a}), 
$\psi_{L}(n,t)=\sum_{k=1}^{3}\lambda_{k,L}E_{k,L}(n,t)$, 
is represented schematically as sketched below 

\begin{figure}[!hbt]
\begin{center}
\epsffile{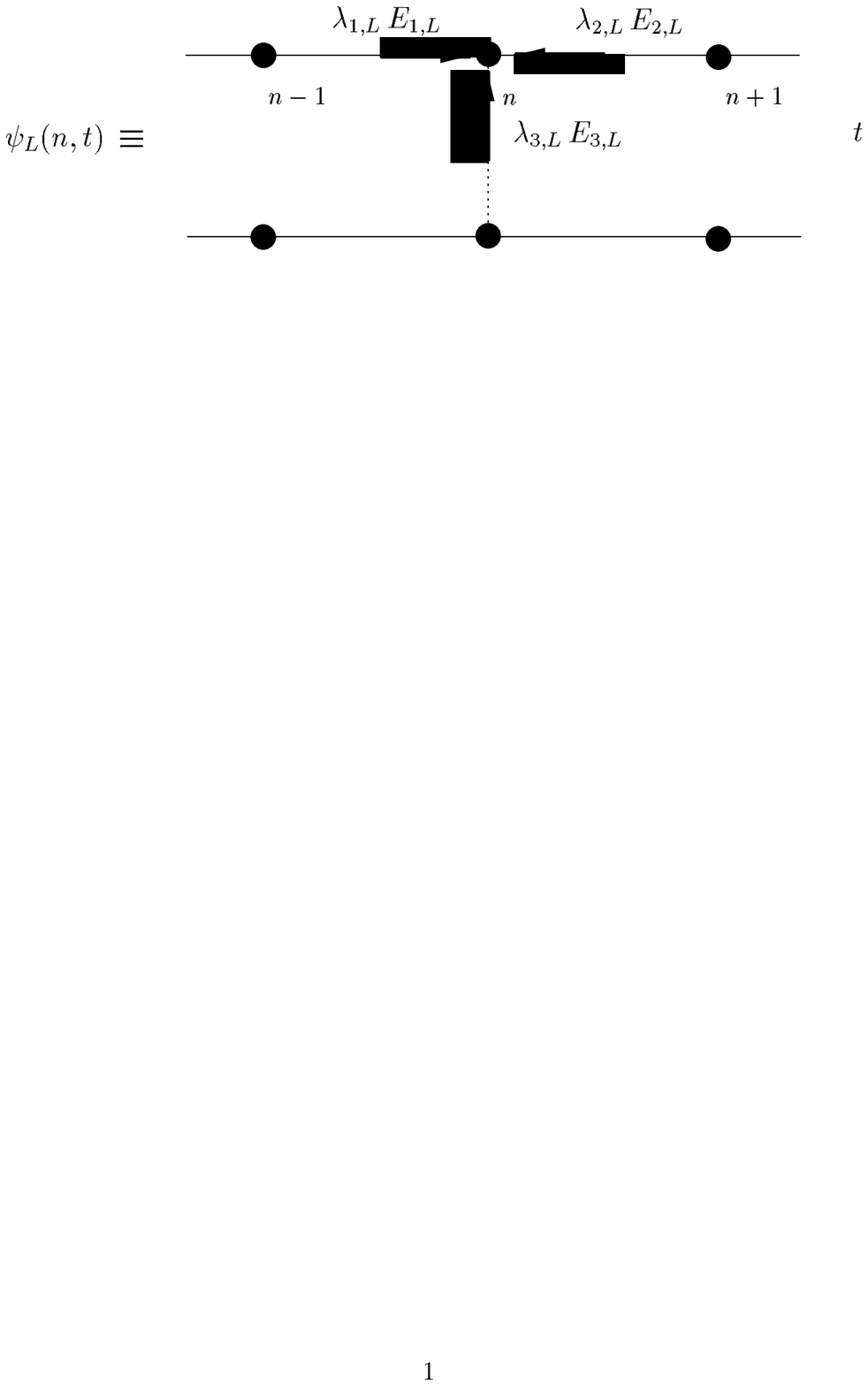}
\end{center}
\end{figure}
whereas equation \refp{eq:psiL_int_1} is sketched 

\begin{figure}[!hbt]
\begin{center}
\epsffile{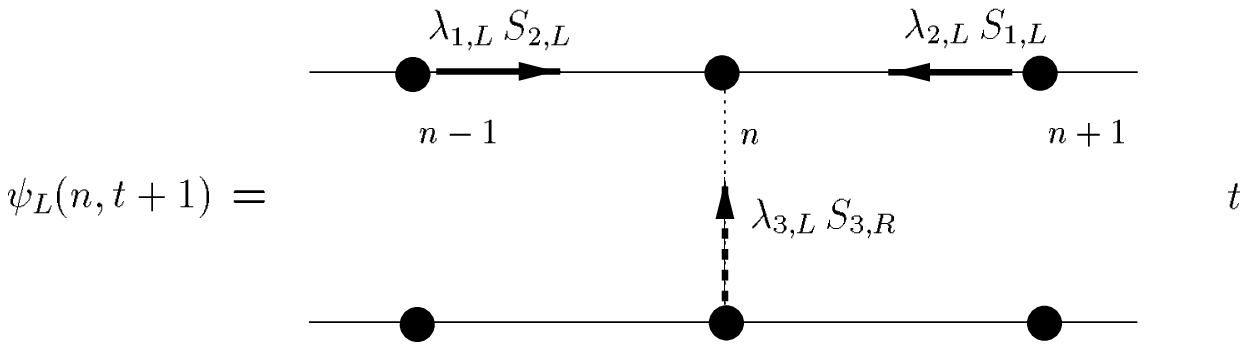}
\end{center}
\end{figure}
\newpage
At time $t$, the outgoing bond currents appearing in equation \refp{eq:psiL_int_1}
were instantaneously scattered by $S_L$ matrices.  
Hence, $S_{1,L}$ and $S_{2,L}$ are linear superpositions 
of $L$ incident currents on nodes $n+1$ and $n-1$ respectively, whereas $S_{3,R}$ is 
scattered by an $S_R$ matrix on node $n$. The resulting
linear superpositions are given by the specific form \refp{eq:outcurrent}. 
Schematically, this scattering process is represented as follows

\begin{figure}[!hbt]
\epsffile{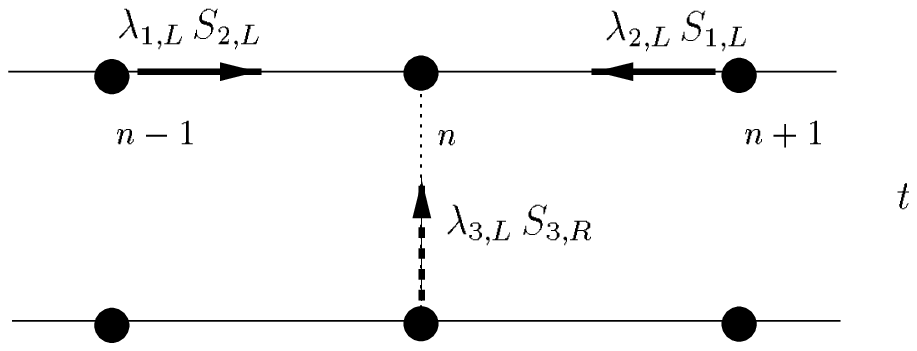}
\end{figure}
\begin{figure}[!hbt]
\begin{center}
\epsffile{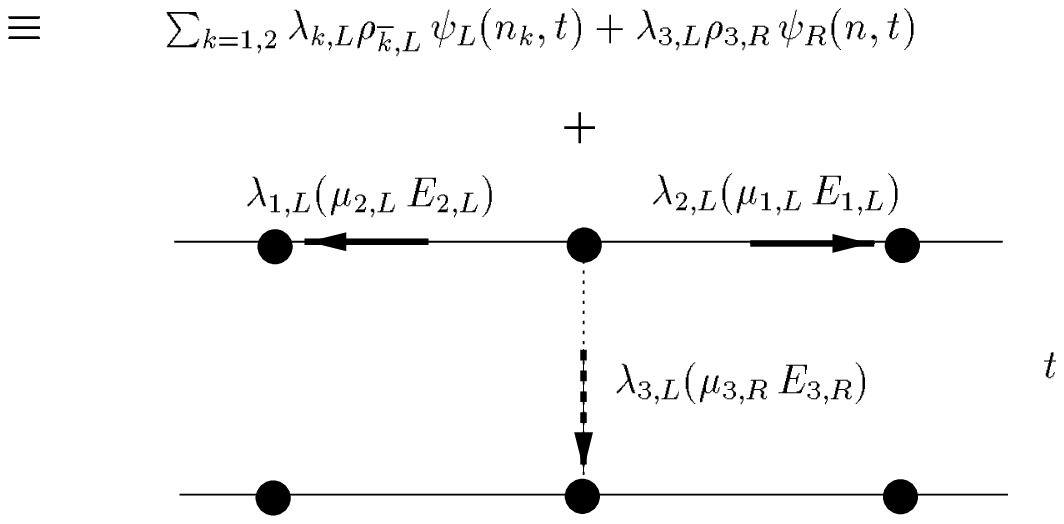}
\end{center}
\end{figure}
\begin{figure}[hbt]
\epsffile{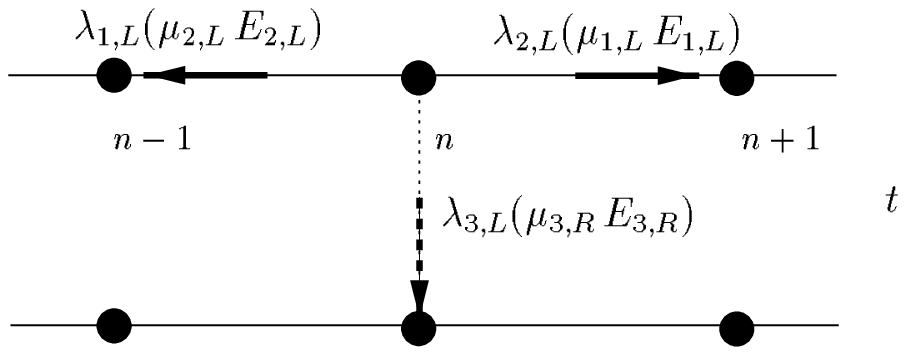}
\centerline{\epsffile{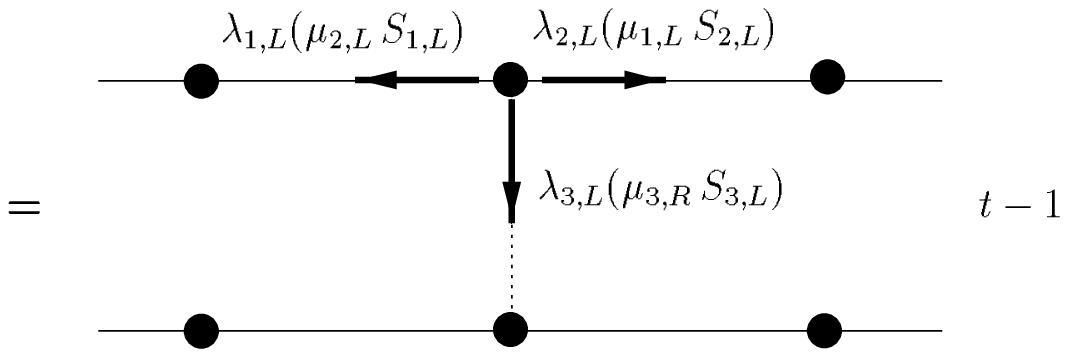}}
\caption{The upper drawing is a sketch of the current term in 
equation~\refp{eq:psiL_int_2} at time $t$, while the lower drawing is a sketch 
of the same currents at time $t-1$ before the propagation step.}
\label{fig:prop_chiral}
\end{figure}
\begin{figure}[hbt]
\centerline{\epsffile{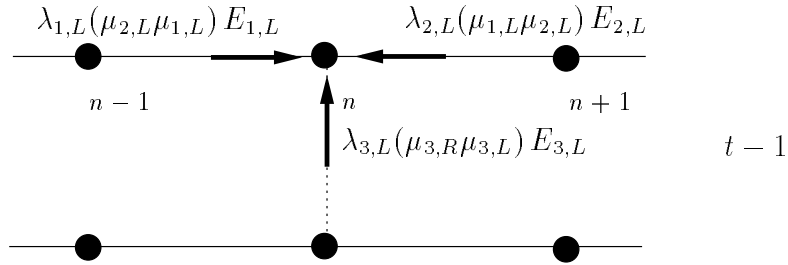}}
\caption{Sketch of the new current term in equation~\refp{eq:psiL_int_4}.}
\label{fig:psiL_3mod}
\end{figure}

which reads analytically

\begin{eqnarray}
\lefteqn{\psi_L(n,t+1) = \sum_{k=1,2} \lambda_{k,L} \rho_{\overline{k},L} 
\, \psi_L(n_k,t) + \lambda_{3,L} \rho_{3,R} \, \psi_R(n,t)} \nonumber \\
& & +\underbrace{\sum_{k=1,2} \lambda_{k,L} \mu_{\overline{k},L} \, 
E_{\overline{k},L}(n_k,t) + \lambda_{3,L} \mu_{3,R} \, 
E_{3,R}(n,t).}_{\mbox{current term}} \label{eq:psiL_int_2}
\end{eqnarray}
According to the propagation step, the current term 
of equation \refp{eq:psiL_int_2}, which is the sum involving only the 
incident currents and not the chiral fields, is deduced from three 
outgoing currents at time $t-1$. Those outgoing currents are all
defined on node $n$, as sketched in Figure \ref{fig:prop_chiral}
\newpage
which translates analytically through the following equation

\begin{equation}
\mbox{current term} \, =
\sum_{k=1,2} \lambda_{k,L} \mu_{\overline{k},L} \, S_{k,L}(n,t-1) +
\lambda_{3,L} \mu_{3,R} \, S_{3,L}(n,t-1). 
\label{eq:psiL_int_3}
\end{equation} 
Again, the outgoing currents appearing in the right hand side of 
equation \refp{eq:psiL_int_3} are deduced from incident currents on the same node according
to the scattering process~\refp{eq:outcurrent}. This leads to the following equation 

\begin{eqnarray}
\lefteqn{\mbox{current term} \, = \,
	\left(\sum_{k=1,2} \lambda_{k,L} \rho_{k,L} \mu_{\overline{k},L} 
        + \lambda_{3,L} \rho_{3,L} \mu_{3,R}\right) \, \psi_L(n,t-1)} \nonumber \\
& &     +\underbrace{ \sum_{k=1,2} \lambda_{k,L} \mu_{\overline{k},L} \mu_{k,L} \, 
	E_{k,L}(n,t-1) + 
	\lambda_{3,L} \mu_{3,R} \mu_{3,L} \, E_{3,L}(n,t-1) }_{\mbox{new current term}}. 
\label{eq:psiL_int_4}
\end{eqnarray}
The new current term in the last equation is schematically depicted in 
Figure~\ref{fig:psiL_3mod}

Plugging the right hand side of equation \refp{eq:psiL_int_4} into 
equation \refp{eq:psiL_int_2} leads to

\begin{eqnarray}
\lefteqn{\psi_L(n,t+1) - \left(
\sum_{k=1,2} \lambda_{k,L} \rho_{k,L} \mu_{\overline{k},L} 
+ \lambda_{3,L} \rho_{3,L} \mu_{3,R}\right) \, \psi_L(n,t-1) \, = } 
\nonumber \\
& &     \sum_{k=1,2} \lambda_{k,L} \rho_{\overline{k},L} \, \psi_L(n_k,t) 
        +\lambda_{3,L} \rho_{3,R} \, \psi_R(n,t) \nonumber \\ 
& &     +\sum_{k=1,2} (\mu_{\overline{k},L} \mu_{k,L}) \lambda_{k,L} \, 
        E_{k,L}(n,t-1) + (\mu_{3,R} \mu_{3,L}) \lambda_{3,L} \, E_{3,L}(n,t-1). 
\label{eq:psiL_int_5}
\end{eqnarray}
Equation~\refp{eq:psiL_int_5} is almost in the closed form we are looking for but still
contains a current term. In order to get an equation which involves 
solely $L,R$ fields without any $L,R$ currents, it is sufficient to impose the following relations 

\begin{equation}
\left\{
\begin{array}{lcl}
\mu_{k,L} \mu_{\overline{k},L} & = & {\mu_L}^2, \esp k=1,2, \\
\\
\mu_{3,R} \mu_{3,L} & = & {\mu_L}^2,
\end{array}
\right. 
\label{eq:condsufL}
\end{equation}
where $\mu_{L}$ is a constant parameter. The relations~\refp{eq:condsufL}
have been chosen in such a way that the last term of equation \refp{eq:psiL_int_5} becomes 
proportional to $\psi_L(n,t-1)$. The insertion of equations~\refp{eq:condsufL} 
into~\refp{eq:psiL_int_5} 
leads to a discrete propagation equation of the type of equation \refp{eq:wave_field_L}~: 
 
\begin{equation}
\psi_L(n,t+1) - {\mu_L}^2\left( 1+\sum_{k=1}^{3} \frac{\lambda_{k,L} 
\rho_{k,L}}{\mu_{k,L}}\right) \, \psi_{L}(n,t-1)  = 
 \sum_{k=1,2} \lambda_{k,L} \rho_{\overline{k},L} \, \psi_{L}(n_k,t) 
        + \lambda_{3,L} \rho_{3,R} \, \psi_{R}(n,t). 
\label{eq:field_L}
\end{equation}
The discrete propagation equation obeyed by the $R$ field is directly deduced from the 
equation for the $L$ field by exchanging all the subscripts $L$ and $R$~:

\begin{equation}
\psi_R(n,t+1) - {\mu_R}^2\left( 1+\sum_{k=1}^{3} \frac{\lambda_{k,R} 
\rho_{k,R}}{\mu_{k,R}}\right) \, \psi_{R}(n,t-1) =
 \sum_{k=1,2} \lambda_{k,R} \rho_{\overline{k},R} \, \psi_{R}(n_k,t) 
        + \lambda_{3,R} \rho_{3,L} \, \psi_{L}(n,t), 
\label{eq:field_R}
\end{equation}
where equation \refp{eq:field_R} has been supplemented by sufficient 
conditions of the type of equations \refp{eq:condsufL} 

\begin{equation}
\left\{
\begin{array}{lcl}
\mu_{k,R} \mu_{\overline{k},R} & = & {\mu_R}^2, \esp k=1,2, \\ 
\\
\mu_{3,L} \mu_{3,R} & = & {\mu_R}^2.
\end{array}
\right.
\label{eq:condsufR}
\end{equation}
The comparison of the second equations in both systems~\refp{eq:condsufL}, \refp{eq:condsufR}
leads to the equality~: ${\mu_L}^2 = {\mu_R}^2$. 
In the rest of this section, we will denote this 
constant by $\mu^2$. Now, if the matrix parameters appearing in  
equations \refp{eq:field_L}, \refp{eq:field_R} are chosen according 
to the following relations

\begin{equation}
\mu^2 \left( 1 + \sum_{k=1}^{3} \frac{\lambda_{k,L} \rho_{k,L}}{\mu_{k,L}}
\right) =  \mu^2 \left( 1 + \sum_{k=1}^{3} 
\frac{\lambda_{k,R} \rho_{k,R}}{\mu_{k,R}} \right) \; = \; 1, 
\label{eq:iddirac_a} 
\end{equation}
then a first-order discretized time derivative appears

\begin{eqnarray}
\psi_L(n,t+1) - \psi_{L}(n,t-1) & = &
\sum_{k=1,2} \lambda_{k,L}\rho_{\overline{k},L} \psi_{L}(n_k,t) 
+ \lambda_{3,L}\rho_{3,R} \, \psi_{R}(n,t), 
\label{eq:field_Lmod1}
\\
\psi_R(n,t+1) - \psi_{R}(n,t-1) & = &
\sum_{k=1,2} \lambda_{k,R}\rho_{\overline{k},R} \psi_{R}(n_k,t) 
+ \lambda_{3,R}\rho_{3,L} \, \psi_{L}(n,t).
\label{eq:field_Rmod1}
\end{eqnarray}
On the same ground, the sums 
$\sum_{k=1,2} \lambda_{k,L,R}\rho_{\overline{k},L,R} \psi_{L,R}(n_k,t)$ 
entering equations \refp{eq:field_Lmod1}, \refp{eq:field_Rmod1} 
can be turned into  first-order discretized spatial derivatives by 
an appropriate choice of the products 
$\lambda_{k,L,R}\rho_{\overline{k},L,R}$, for $k=1,2$. 
With this assumption, 
the two coupled discretized wave propagation equations \refp{eq:field_Lmod1},
\refp{eq:field_Rmod1} can be identified with the Dirac equation in 
the Majorana-Weyl representation. However, we don't need this last assumption
at this step of our derivation. In the next section we
show that implementing some suitable symmetries on the scattering process
constrains even more the $L,R$ matrix coefficients and turns the terms
$\sum_{k=1,2} \lambda_{k,L,R}\rho_{\overline{k},L,R} \psi_{L,R}(n_k,t)$
into first-order discretized spatial derivatives.

\subsection{Symmetries of the scattering process}~\label{symet_maj}
Our goal is now to prove that a few well chosen assumptions on the
symmetries of the scattering process leads to the unique determination
of the $L,R$ scattering matrices and to the discrete
propagation equations \refp{eq:field_Lmod1}, \refp{eq:field_Rmod1} in a form
very close to the Dirac equation \refp{eq:contdirac}. The symmetries are implemented both 
on the currents and on the fields. In this model, the scattering process is 
described by matrices which are unitary due to
time-reversal invariance and reciprocity. This property insures the local 
conservation, at each time step, of the ``flux'' of the $L,R$ currents~: 
$(\module{S_{1}}^2 + \module{S_{2}}^2 +
\module{S_{3}}^2)_{L,R} = (\module{E_{1}}^2 + \module{E_{2}}^2 +
\module{E_{3}}^2)_{L,R}$,   
which is a stability condition on the dynamics of the $L,R$ fields 
at large times. Finally, the last symmetry required to complete the construction
is reflection symmetry ($x \mapsto -x$) which is known to connect
the two chiral fields.

\subsubsection{Time-reversal invariance}
\begin{figure}[hbt]
\centerline{\epsffile{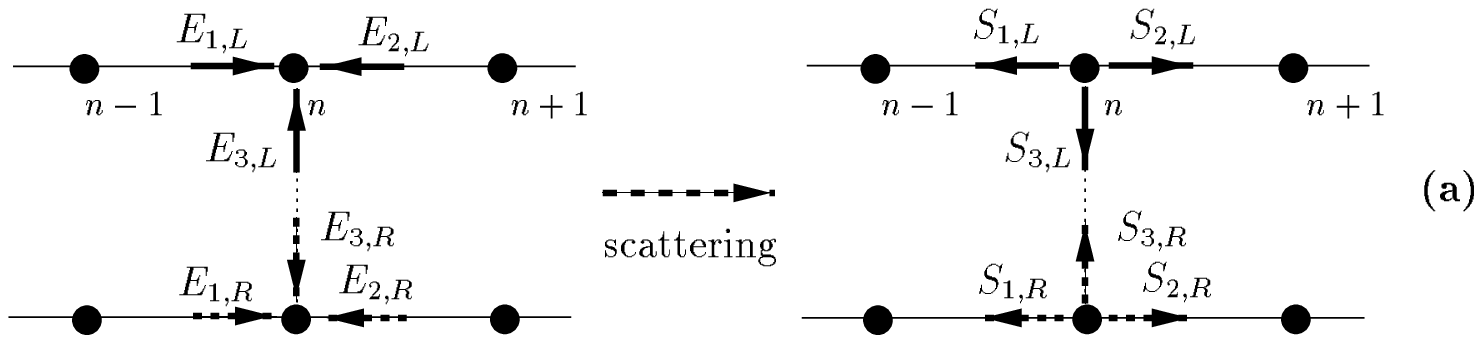}}
\centerline{\epsffile{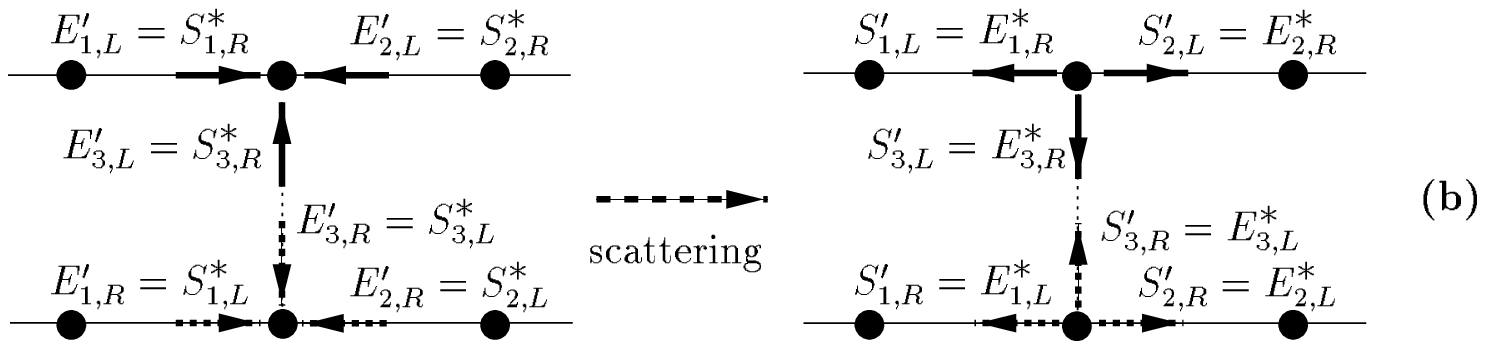}}
\caption{Time-reversal invariance of the scattering process.
The scattering process for the $L,R$ currents depicted in {\bf (a)} is equivalent
to the scattering process deduced from time-reversal symmetry {\bf (b)}.}
\label{fig:it_chiral}
\end{figure}

Time-reversal invariance for the currents can be 
formulated by stating that the two scattering processes, sketched in 
Figure \ref{fig:it_chiral}, are equivalent under a time-reversal 
transformation 

Note that the $L$ and $R$ currents are exchanged in this transformation. This 
corresponds to the exchange of the spinor components $\psi_L$ and $\psi_R$ when
considering the time-reversal transformation in the Dirac equation. This leads to 
the following relation between the scattering matrices 
 
\begin{equation}
S_{L} = ({S_{R}}^{\ast})^{-1}. \label{eq:it_matrix}
\end{equation}
According to the form \refp{def:chiral_scatt_a}, \refp{def:chiral_scatt_b}, 
\refp{def:chiral_scatt_c} of the $L,R$ scattering matrices, equation~\refp{eq:it_matrix}
leads to the following constraints on the $L,R$ matrix coefficients

\begin{equation}
\left\{
\begin{array}{lclclcllc}
\mu_{k,L} \displaystyle{\frac{\rho_{k,R}^{\ast}}{\rho_{k,L}}} & = & C, && 
\mu_{k,R} \displaystyle{\frac{\lambda_{k,L}^{\ast}}{\lambda_{k,R}}} & = 
& C', &&\\
&&&&&& && \mbox{for} \esp k=1,2,3,\\
\mu_{k,L} \displaystyle{\frac{\lambda_{k,R}^{\ast}}{\lambda_{k,L}}} & = &
\displaystyle{\frac{1}{C'^{\ast}}}, && 
\mu_{k,R} \displaystyle{\frac{\rho_{k,L}^{\ast}}{\rho_{k,R}}} & = & 
\displaystyle{\frac{1}{C^{\ast}}}, &&
\end{array}
\right.
\label{eq:it_matrixcoef0}
\end{equation}
where $C,C'$ are two constants which obey the condition
$C + C'^{\ast} = \sum_{k=1}^{3} \lambda_{k,L}\rho_{k,R}^{\ast}$. 
Equations~\refp{eq:it_matrixcoef0} allow to express the $R$ matrix 
coefficients in terms of those of the $L$ matrix.

We must also consider time-reversal invariance for the chiral fields $\psi_L$ and
$\psi_R$. For its implementation, it is convenient to introduce the scalar 
field $\psi^S$ constructed as a linear superposition of outgoing currents
 
\begin{equation}
\psi^{S}(n,t) \equiv \sum_{k=1}^{3} \kappa_{k} S_{k}(n,t),
\label{def:outfield}
\end{equation}
where the coefficients $\kappa_{k}$ are complex numbers. Up to now the
scalar fields $\psi^E_L, \psi^E_R$, considered so far were linear superpositions
of incident currents (equations~\refp{def:chiral_comp_a},
\refp{def:chiral_comp_b}). Owing to the linearity
of the scattering process \refp{eq:scatt_majL},
\refp{eq:scatt_majR} there always exist
numbers $\kappa_{k}$ such that the fields $\psi^S$ obey the same 
propagation equations as those satisfied by $\psi^E$, namely equations
\refp{eq:field_Lmod1}, \refp{eq:field_Rmod1}. 
By taking into account the special form~\refp{eq:outcurrent} of the 
outgoing currents, equation~\refp{def:outfield} for the $L$ field becomes

\[
\psi^{S}_L(n,t) = \left(\sum_{k=1}^{3} \kappa_{k,L}\rho_{k,L}\right) \psi^{E}_L(n,t)
+ \sum_{k=1}^{3} \kappa_{k,L}\mu_{k,L} E_{k,L}(n,t).
\]
It is easy to show that the condition $\kappa_{k,L}\mu_{k,L} = \kappa_L\lambda_{k,L}$, where
$\kappa_{L}$ doesn't depend on the subscript $k$ and is given by
$\kappa^{-1}_L = 1 + \sum_{k=1}^{3} \lambda_{k,L}\rho_{k,L}/\mu_{k,L}$, leads to the
equality between the outgoing field and the incident field~:
$\psi^S_L = \psi^E_L$. Finally, using equations~\refp{eq:iddirac_a},
$\psi^S_L$ reads
 
\begin{equation}
\psi^S_L(n,t) \equiv \sum_{k=1}^{3} \kappa_{k,L}S_{k,L}(n,t) =  \mu^{2} \sum_{k=1}^{3} \frac{\lambda_{k,L}}{\mu_{k,L}} S_{k,L}(n,t).
\label{eq:outfield1}
\end{equation}
Similarly, one finds
 
\begin{equation}
\psi^S_R(n,t) \equiv \sum_{k=1}^{3} \kappa_{k,R}S_{k,R}(n,t) =  \mu^{2} \sum_{k=1}^{3} \frac{\lambda_{k,R}}{\mu_{k,R}} S_{k,R}(n,t).
\label{eq:outfield1R}
\end{equation}
It is natural to implement time-reversal invariance on the chiral fields
by the following conditions
 
\begin{eqnarray}
\psi_L^S(n,t) & = & {\psi_R^E}^{\ast}(n,t) \label{def:it_fields_a},\\
\psi_R^S(n,t) & = & {\psi_L^E}^{\ast}(n,t) \label{def:it_fields_b}.
\end{eqnarray}
Here, we have explicitly distinguished the fields constructed on the
outgoing currents, $\psi_{L,R}^S$, and the fields constructed on the
incident currents, $\psi_{L,R}^E$. Moreover the complex conjugate fields 
have been considered in the inverse process. Now if we write down the currents
in equations~\refp{def:it_fields_a}, \refp{def:it_fields_b} according to
the definitions of the fields
\refp{def:chiral_comp_a}, \refp{def:chiral_comp_b}, \refp{def:outfield}
the time-reversal invariance of the 
scattering process (Fig.\ \ref{fig:it_chiral}) leads to the following relations

\begin{equation}
\kappa_{k,L} = \lambda_{k,R}^{\ast}, \esp \kappa_{k,R} = 
\lambda_{k,L}^{\ast} \esp\esp \mbox{for} \esp k=1,2,3.
\label{eq:it_lambda}
\end{equation}
Then, using~\refp{eq:condsufL}, \refp{eq:condsufR} 
with $\mu_{L}^2=\mu_{R}^2=\mu^2$, the last equation~\refp{eq:it_lambda} gives

\begin{equation}
\left\{
\begin{array}{ccllcccc}
\mu_{k,R} \mu_{k,L}^{\ast} & = & 1, & k=1,2,3, && \mbox{and} &&
\module{\mu} \; = \; 1, \\ 
\\
\mu_{\overline{k},L} & = & \mu^2 \mu_{k,R}^{\ast}, & k=1,2. &&&&
\end{array}
\right.
\label{eq:module_mu}
\end{equation}
Equations~\refp{eq:it_matrixcoef0}, \refp{eq:it_lambda} and \refp{eq:module_mu} lead to~: 
$C'=\mu^2$. Lastly, the relations~\refp{eq:it_matrixcoef0} read

\begin{equation}
\left\{
\begin{array}{lclclcllc}
\mu_{k,L} \displaystyle{\frac{\lambda_{k,R}^{\ast}}{\lambda_{k,L}}} & = & \mu^2, && 
\mu_{k,R} \displaystyle{\frac{\lambda_{k,L}^{\ast}}{\lambda_{k,R}}} & = & \mu^2, &&\\
&&&&&& && \mbox{for} \esp k=1,2,3,\\
\mu_{k,L} \displaystyle{\frac{\rho_{k,R}^{\ast}}{\rho_{k,L}}} & = & C, && 
\mu_{k,R} \displaystyle{\frac{\rho_{k,L}^{\ast}}{\rho_{k,R}}} & = & \displaystyle{\frac{1}{C^{\ast}}}, &&
\end{array}
\right.
\label{eq:it_matrixcoef}
\end{equation} 
with $C+(\mu^{\ast})^2=\sum_{k=1}^{3} \lambda_{k,L}\rho^{\ast}_{k,R}$.

\subsubsection{Reciprocity principle} 
\begin{figure}[hbt]
\centerline{\epsffile{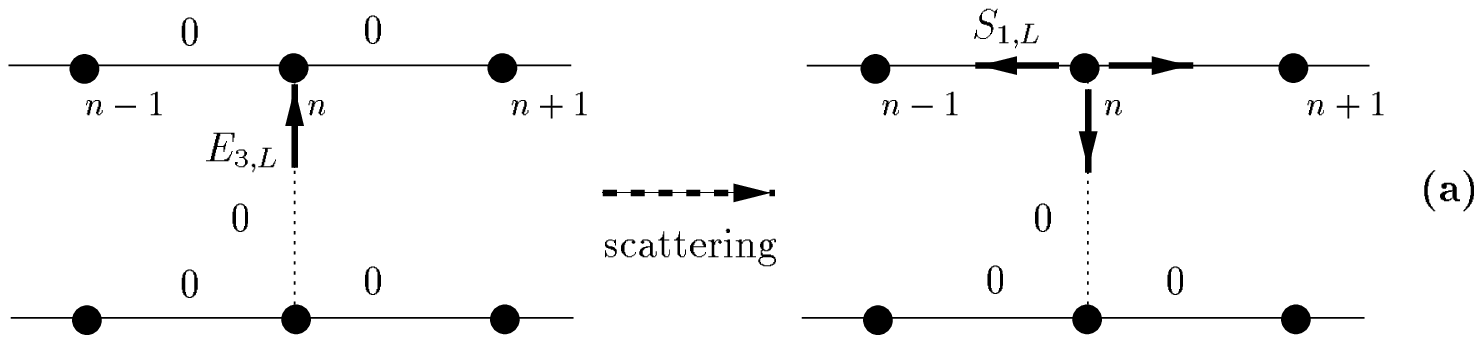}}
\centerline{\epsffile{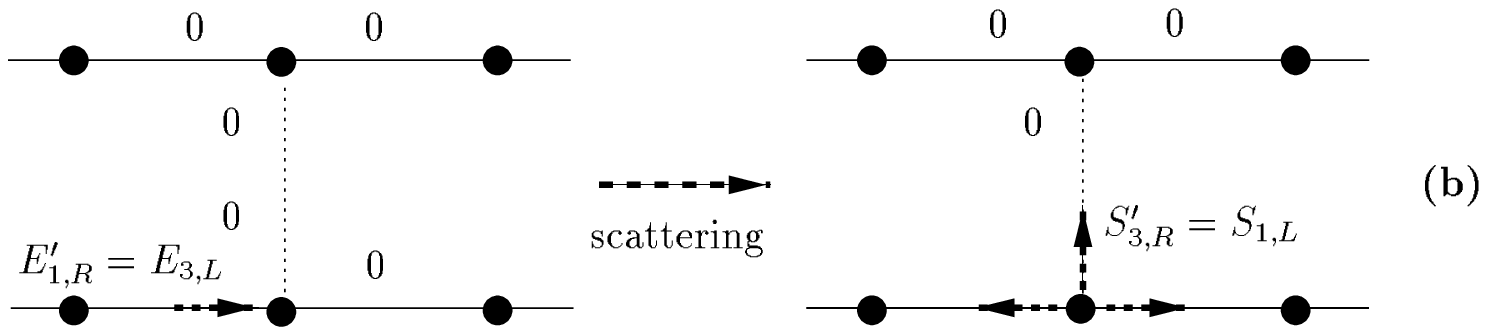}}
\caption{Reciprocity principle of the scattering process. The scattering channel 
$3\rightarrow 1$ for the $L$ currents {\bf (a)} is equivalent to the scattering
channel $1\rightarrow 3$ for the $R$ currents {\bf (b)}.}
\label{fig:rec_chiral}
\end{figure}

The reciprocity principle states that the response of a system at a point $\mbox{\boldmath{$r'$}}$
due to an excitation at point $\mbox{\boldmath{$r$}}$ is identical to the reciprocal process, 
namely the response of the system at point $\mbox{\boldmath{$r$}}$ due to an excitation at point 
$\mbox{\boldmath{$r'$}}$. In our current model, the application of this principle to the 
scattering process means that any scattering channel is identical to its reciprocal channel. 
The definition of a scattering channel is illustrated in
Figure~\ref{fig:rec_chiral} {\bf (a)}. An incident $L$ current at node $n$, $E_{3,L}$ in 
Figure~\ref{fig:rec_chiral} {\bf (a)}, is scattered in three outgoing currents $S_{i,L}$.
A scattering channel is any of the three elementary processes, $E_{3,L}\rightarrow S_{1,L}$,
$E_{3,L}\rightarrow S_{2,L}$ and $E_{3,L}\rightarrow S_{3,L}$. Let us consider the first channel
$E_{3,L}\rightarrow S_{1,L}$. The reciprocity principle states that this scattering channel is 
equivalent to the reciprocal channel for the $R$ current illustrated in 
Figure~\ref{fig:rec_chiral} {\bf (b)}. Roughly speaking, each scattering channel for $L$ 
currents in one direction is equivalent to the same scattering channel for $R$ currents 
in the opposite direction. As for the time-reversed processes considered in the previous 
section, note that the $L$ and $R$ currents are exchanged in the two reciprocal processes. 
This exchange of $L$ and $R$ currents is suggested by the exchange of the
spinor components $\psi_L$ and $\psi_R$ in the symmetries underlying the Dirac equation.  

Finally, reciprocity as defined above leads to the 
following relation between the scattering matrices

\begin{equation}
S_L = {}^T S_R, \label{eq:rec_matrix}
\end{equation}
and to the following constraints on the $L,R$ matrix coefficients

\begin{equation}
\left\{
\begin{array}{lcllc}
\rho_{k,L} & = & \gamma \lambda_{k,R}, && \\
\rho_{k,R} & = & \gamma \lambda_{k,L}, && \mbox{for} \esp k=1,2,3, \\
\mu_{k,L} & = & \mu_{k,R}, &&
\end{array}
\right.
\label{eq:rec_matrixcoef}
\end{equation}
where $\gamma$ is a complex number.

\subsubsection{Unitarity of the scattering process}~\label{sym_maj_unitarity} 
The combination of  equations~\refp{eq:it_matrix} and \refp{eq:rec_matrix} 
shows that $S_L$ and $S_R$ are unitary matrices, each one satisfying~:

\begin{equation}
S S^{\dag} = \mbox{Id},
\label{def:unitarity}
\end{equation}
where $\mbox{Id}$ denotes the identity matrix. Unitarity which 
is expressed by combining together equations~\refp{eq:module_mu}, 
\refp{eq:it_matrixcoef} and \refp{eq:rec_matrixcoef} leads to
\begin{equation}
\left\{
\begin{array}{ccllc}
\module{\mu_{k,L}} & = & \module{\mu_{k,R}} = 1, && \\
\\
\mu_{k,L} \displaystyle{\frac{\lambda_{k,L}^{\ast}}{\rho_{k,L}}} & = & 
\displaystyle{\frac{C}{\gamma^{\ast}}}, && \mbox{for} \esp k=1,2,3, \\
\\
\mu_{k,R} \displaystyle{\frac{\lambda_{k,R}^{\ast}}{\rho_{k,R}}} & = &
\displaystyle{\frac{1}{(C\gamma)^{\ast}}}, &&
\end{array}
\right.
\label{eq:un_matrixcoef}
\end{equation}
where $C/\gamma + (C/\gamma)^{\ast} = -\Lambda$, with 
$\Lambda \equiv \sum_{k=1,2,3}\module{\lambda_{k,L}}^2=$ $\sum_{k=1,2,3}\module{\lambda_{k,R}}^2$. 
Finally, some manipulations using equations~\refp{eq:iddirac_a}, \refp{eq:rec_matrixcoef} and 
\refp{eq:un_matrixcoef} yields the parametrization of the $L,R$ matrix coefficients in terms of 
$\lambda_{k,L}, \mu_{k,L}$ and $\mu$
 
\begin{equation}
\left\{
\begin{array}{ccllc}
\rho_{k,L} & = & \displaystyle{\frac{\mu^{\ast}-\mu}{\Lambda}} 
\left(\displaystyle{\frac{\mu_{k,L}}{\mu}}\right) \lambda_{k,L}^{\ast}, && \\
\\
\lambda_{k,R} & = & \displaystyle{\frac{1}{\mu}} 
\left(\displaystyle{\frac{\mu_{k,L}}{\mu}}\right) \lambda_{k,L}^{\ast}, &&
\mbox{for} \esp k=1,2,3,\\
\\
\rho_{k,R} & = & \displaystyle{\frac{\mu^{\ast}-\mu}{\Lambda}}\mu \lambda_{k,L}. &&
\end{array}
\right.
\label{eq:parametriz1}
\end{equation}
Using this parametrization and choosing $\mu_{3,L}=\mu_{3,R}=\mu$ from equations
\refp{eq:condsufL}, \refp{eq:rec_matrixcoef} the scattering matrices $S_L$
and $S_R$, (\refp{def:chiral_scatt_a}, \refp{def:chiral_scatt_b}, 
\refp{def:chiral_scatt_c}) take the form

\begin{eqnarray}
s_{kl,L} & = & \displaystyle{\frac{\mu^{\ast}-\mu}{\Lambda}} 
\left(\displaystyle{\frac{\mu_{k,L}}{\mu}}\right) \lambda_{k,L}^{\ast}
\lambda_{l,L} + \mu_{k,L}\delta_{kl},
\label{def:chiral_scattmod1} \\
\nonumber\\
s_{ij,R} & = & \displaystyle{\frac{\mu^{\ast}-\mu}{\Lambda}} 
\left(\displaystyle{\frac{\mu_{j,L}}{\mu}}\right) \lambda_{i,L}
\lambda_{j,L}^{\ast} + \mu_{i,L}\delta_{ij},
\label{def:chiral_scattmod2}
\end{eqnarray}
and the wave propagation equations~\refp{eq:field_Lmod1}, \refp{eq:field_Rmod1} become

\begin{eqnarray}
\lefteqn{ \psi_L(n,t+1) - \psi_{L}(n,t-1) \: =} \nonumber \\
&&	\displaystyle{\frac{\mu^{\ast}-\mu}{\Lambda}}
	\left[\left(\displaystyle{\frac{\mu_{1,L}}{\mu}}\right)
	\lambda_{1,L}^{\ast}\lambda_{2,L} \, 
	\psi_{L}(n+1,t) + 
	\left(\displaystyle{\frac{\mu_{2,L}}{\mu}}\right)
	\lambda_{1,L}\lambda_{2,L}^{\ast} \, 
	\psi_{L}(n-1,t)\right] + 
	\gamma\lambda_{3,L}^{2} \, \psi_{R}(n,t), \label{eq:field_Lmod2} \\
&& \nonumber \\
\lefteqn{ \psi_R(n,t+1) - \psi_{R}(n,t-1) \: =} \nonumber \\
&&	\displaystyle{\frac{\mu^{\ast}-\mu}{\Lambda}} 
	\left[\left(\displaystyle{\frac{\mu_{2,L}}{\mu}}\right) 
	\lambda_{1,L}\lambda_{2,L}^{\ast} \, \psi_{R}(n+1,t) 
	+\left(\displaystyle{\frac{\mu_{1,L}}{\mu}}\right) 
	\lambda_{1,L}^{\ast}\lambda_{2,L} \, \psi_{R}(n-1,t)\right] \nonumber \\
&&	+\left(\frac{\mu^{\ast}-\mu}{\Lambda}\right)^{2}
	\frac{{\lambda_{3,L}^{\ast}}^{2}}{\gamma} \, \psi_{L}(n,t).
	\label{eq:field_Rmod2}
\end{eqnarray}

\subsubsection{Reflection invariance}~\label{sym_maj_parity}
\begin{figure}[hbt]
\centerline{\epsffile{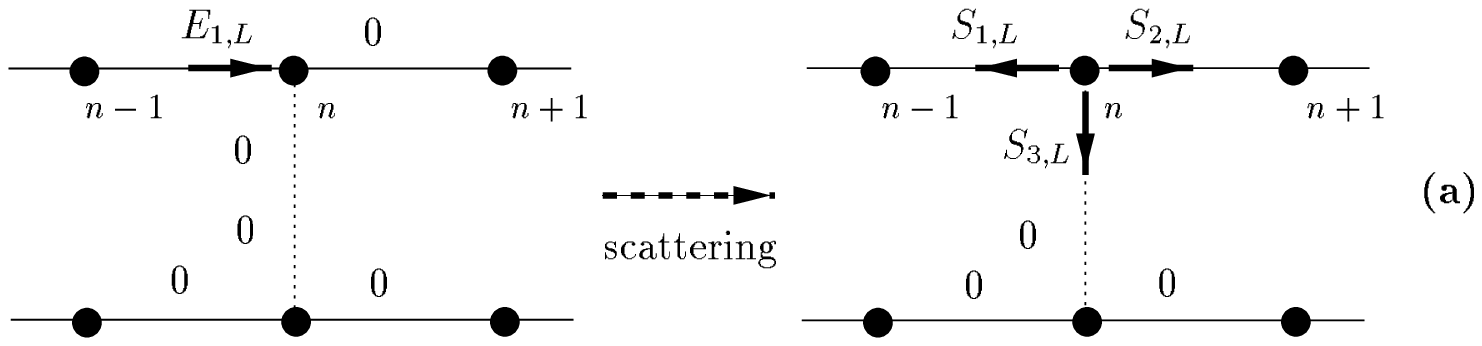}}
\centerline{\epsffile{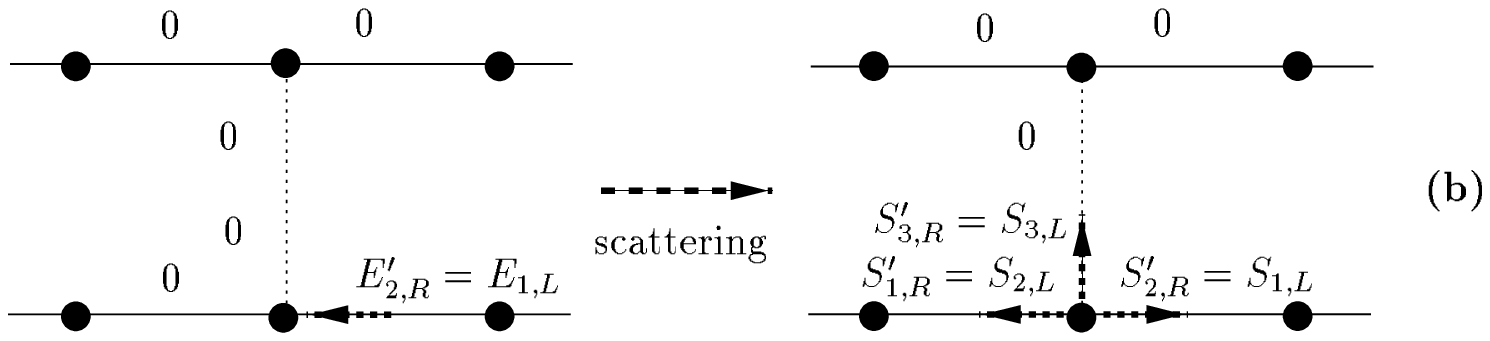}}
\caption{Reflection invariance of the scattering process for the left bond current. The scattering 
process for the $L$ current $E_{1,L}$ {\bf (a)} is equivalent to the scattering process
after a reflection symmetry for the $R$ current $E'_{2,R}$ {\bf (b)}.}
\label{fig:is_chiral1}
\end{figure}

\begin{figure}[hbt]
\centerline{\epsffile{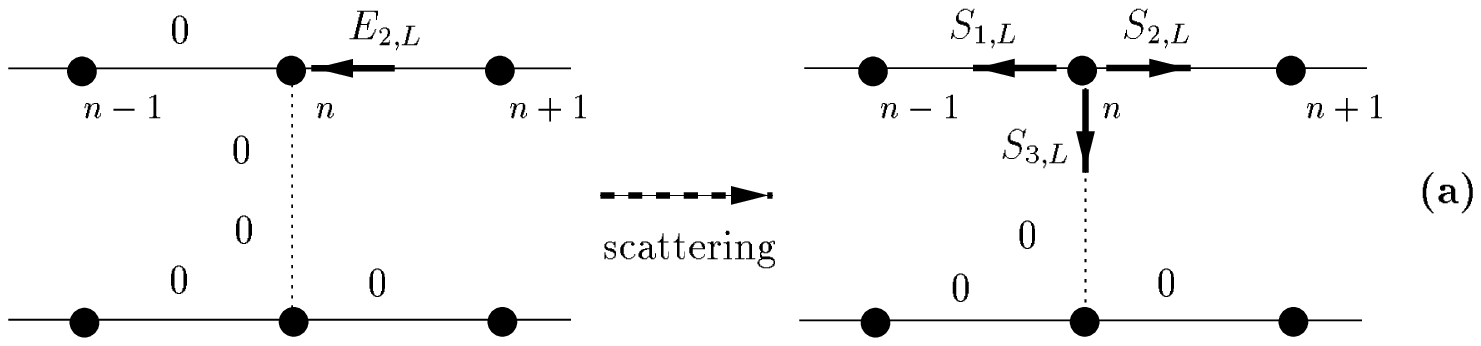}}
\centerline{\epsffile{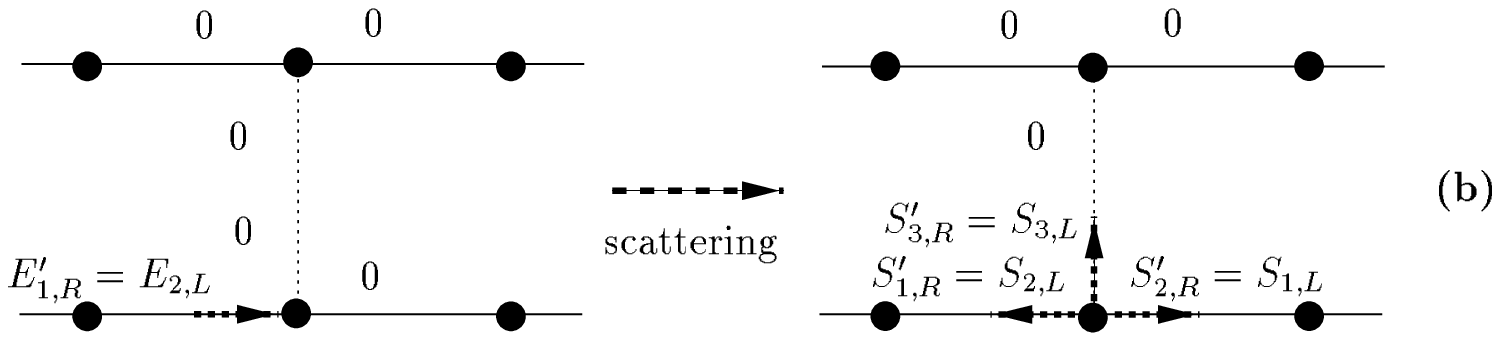}}
\caption{Reflection invariance of the scattering process for the right bond current. The scattering 
process for the $L$ current $E_{2,L}$ {\bf (a)} is equivalent to the scattering process
after a reflection symmetry for the $R$ current $E'_{1,R}$ {\bf (b)}.}
\label{fig:is_chiral2}
\end{figure}

\begin{figure}[hbt]
\centerline{\epsffile{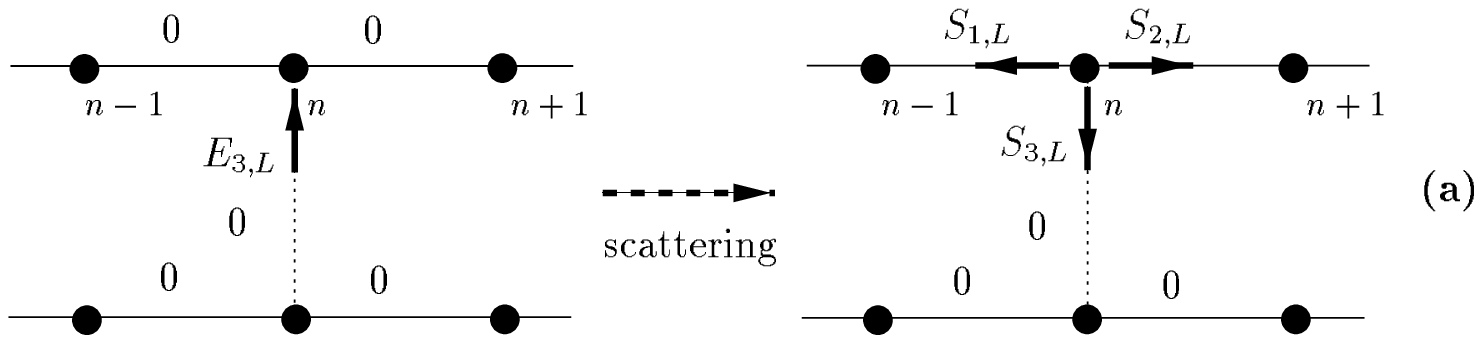}}
\centerline{\epsffile{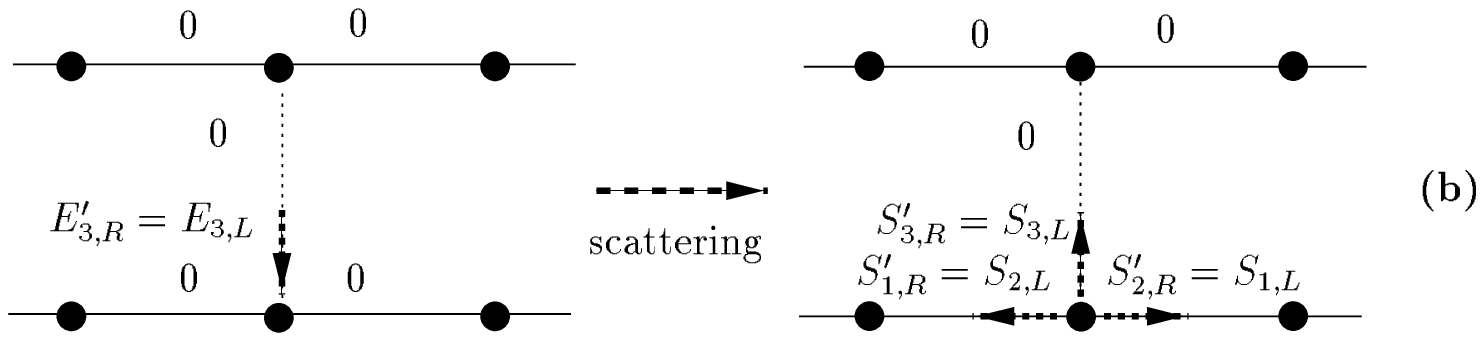}}
\caption{Reflection invariance of the scattering process for the node current. The scattering 
process for $E_{3,L}$ {\bf (a)} is equivalent to the scattering process
after a reflection symmetry for $E'_{3,R}$ {\bf (b)}.}
\label{fig:is_chiral3}
\end{figure}

Reflection invariance states that the Dirac equation~\refp{eq:contdirac} is invariant 
by the simultaneous exchange of $\psi_L$ into $\psi_R$ and $x$ into $-x$. We 
transpose this property to the scattering process of the currents. Reflection invariance 
of the scattering process can be decomposed in three elementary steps sketched in 
Figures~\ref{fig:is_chiral1}, \ref{fig:is_chiral2} and \ref{fig:is_chiral3}.
This symmetry is formulated by stating that the two scattering processes 
represented in figures {\bf (a)} and {\bf (b)} respectively, are equivalent
under a reflection

For example, the equivalence of the scattering processes {\bf (a)} and {\bf (b)}
depicted in Figure \ref{fig:is_chiral1} leads to the following relations between the
matrix elements~: $s_{kl,L} = s_{\overline{k}\, \overline{l},R}, k=1,2,3, \; \mbox{and}
\; l=1$. Similar relations between $s_{kl,L}$ and $s_{\overline{k}\, \overline{l},R}$ hold
from Figures~\ref{fig:is_chiral2} and \ref{fig:is_chiral3}. Finally, the constraints 
imposed by the reflection invariance on the scattering coefficients are summed 
up by the equation

\begin{equation}
s_{kl,L} = s_{\overline{k}\, \overline{l},R}, \esp \mbox{for} \esp k,l=1,2,3.
\label{eq:is_matrixcoef}
\end{equation}
Taking into account the form of the $L$ and $R$ matrix elements 
(\refp{def:chiral_scattmod1}, \refp{def:chiral_scattmod2}), we deduce 
from~\refp{eq:is_matrixcoef}  

\begin{equation}
\left\{
\begin{array}{ccl}
\mu_{1,L} & = & \mu_{2,L}, \\
\\
\displaystyle{\frac{\module{\lambda_{2,L}}}{\module{\lambda_{1,L}}}} & = & 1, \\
\\
\left(\displaystyle{\frac{\lambda_{3,L}}{\lambda_{1,L}}}\right)^{\ast}
\displaystyle{\frac{\lambda_{1,L}}{\lambda_{3,L}}} & = & \displaystyle{
\frac{\mu_{1,L}}{\mu}}\left(\displaystyle{\frac{\lambda_{2,L}}{\lambda_{1,L}}}\right)^{\ast},
\end{array}
\right.
\label{eq:is_matrixcoef1}
\end{equation} 
where, using~\refp{eq:module_mu} and the third equation of system~\refp{eq:rec_matrixcoef},
$\mu_{1,L}$ and $\mu_{2,L}$ can be written

\begin{equation}
\mu_{1,L}=\mu_{2,L}=\varepsilon_{\mu}\mu, \esp \varepsilon_{\mu}=\pm 1.
\label{eq:is_matrixcoef2}
\end{equation}
An additional constraint arises by requiring the reflection invariance of  
the propagation equations~\refp{eq:field_Lmod2}, \refp{eq:field_Rmod2} satisfied by 
the chiral fields. If we plug the last result~\refp{eq:is_matrixcoef2} in those two equations,
the $L,R$ fields are parity invariant provided the last two terms of the
r.h.s. of equations~\refp{eq:field_Lmod2} and \refp{eq:field_Rmod2} are equal, which implies

\begin{equation}
\gamma = \varepsilon_{\gamma}\left(\frac{\mu^{\ast}-\mu}{\Lambda_{L}}\right)
\left(\frac{\lambda_{3,L}^{\ast}}{\lambda_{3,L}}\right), \esp \varepsilon_{\gamma}=\pm 1.
\label{eq:is_matrixcoef3}
\end{equation} 
Using~\refp{eq:is_matrixcoef1} and~\refp{eq:is_matrixcoef3}, we obtain the phase of $\lambda_{3,L}$

\begin{equation}
\frac{\lambda_{3,L}^{\ast}}{\lambda_{3,L}} = \epsilon_\gamma \mu.
\label{eq:phase_gamma}
\end{equation}

\subsection{Identification with the Dirac equation and exact form of the 
scattering matrices}~\label{scatt_maj}
In this section, we show that the propagation equations~\refp{eq:field_Lmod2},
\refp{eq:field_Rmod2} take either the form of the Dirac equation
in the Majorana-Weyl representation or can be identified with two coupled Schr\"odinger 
equations (see Appendix). We first note that the propagation 
equations~\refp{eq:field_Lmod2}, \refp{eq:field_Rmod2} depend only on the ratios 

\begin{equation}
\frac{\lambda_{2,L}}{\lambda_{1,L}}, \esp \mbox{and} \esp 
\nu\equiv\frac{\lambda_{3,L}}{\lambda_{1,L}}.
\label{eq:ratio}
\end{equation}
This is not surprising since the problem is linear which supposes that the chiral fields
are defined up to a multiplicative constant~: $\lambda_{1,L}$. 
Furthermore, according to the equality between the modulus 
of $\lambda_{2,L}$ and $\lambda_{1,L}$ (equation~\refp{eq:is_matrixcoef1}) and to the definition 
of $\Lambda$, those two equations can be recast in the form

\begin{eqnarray}
\lefteqn{ \psi_L(n,t+1) - \psi_{L}(n,t-1) \: =} \nonumber \\
&&	\varepsilon_{\mu}\frac{(\mu^{\ast}-\mu)}{2+{\module{\nu}}^{2}}
	\frac{\lambda_{2,L}}{\lambda_{1,L}}\left[\psi_{L}(n+1,t) + 
	\left(\frac{\lambda_{2,L}}{\lambda_{1,L}}\right)^{\ast}\frac{\lambda_{1,L}}{\lambda_{2,L}} 
	\, \psi_{L}(n-1,t)\right] + 
	\varepsilon_{\gamma}\left(\frac{(\mu^{\ast}-\mu){\module{\nu}}^{2}}
	{2+{\module{\nu}}^{2}}\right) \, \psi_{R}(n,t), 
\label{eq:field_Lmod3} \\
&& \nonumber \\
\lefteqn{ \psi_R(n,t+1) - \psi_{R}(n,t-1) \: = } \nonumber \\
&&	\varepsilon_{\mu}\frac{(\mu^{\ast}-\mu)}{2+{\module{\nu}}^{2}}
	\left(\frac{\lambda_{2,L}}{\lambda_{1,L}}\right)^{\ast}
	\left[\psi_{R}(n+1,t) + 
	\frac{\lambda_{2,L}}{\lambda_{1,L}}\left(\frac{\lambda_{1,L}}{\lambda_{2,L}}\right)^{\ast} 
	\, \psi_{R}(n-1,t)\right] \nonumber \\
&&	+\varepsilon_{\gamma}\left(\frac{(\mu^{\ast}-\mu){\module{\nu}}^{2}}
	{2+{\module{\nu}}^{2}}\right) \, \psi_{L}(n,t).
\label{eq:field_Rmod3}
\end{eqnarray}
If we restrict the ratio $(\lambda_{2,L}/\lambda_{1,L})^{\ast}/(\lambda_{2,L}/\lambda_{1,L})$ to 
be equal to $\pm 1$, i.e. $\lambda_{2,L}/\lambda_{1,L}$ is either a real number or a purely 
imaginary number,  then the r.h.s. of the propagation equations~\refp{eq:field_Lmod3}, 
\refp{eq:field_Rmod3} displays a second order discretized spatial derivative or a first-order 
discretized spatial derivative, respectively. We postpone the case where 
$\lambda_{2,L}/\lambda_{1,L}$ is a real number 
to the appendix and concentrate on the other alternative where $\lambda_{2,L}/\lambda_{1,L}$ 
is a purely imaginary number. Combining 
$(\lambda_{2,L}/\lambda_{1,L})^{\ast}=-(\lambda_{2,L}/\lambda_{1,L})$ with 
$\module{\lambda_{2,L}/\lambda_{1,L}}=1$ (Eq.~\refp{eq:is_matrixcoef1}) leads to 
$\lambda_{2,L}/\lambda_{1,L}=i\varepsilon_{\lambda}$, where $\varepsilon_{\lambda}=\pm1$. 
Writing $\mu=\exp(i\theta)$ (see Eq.~\refp{eq:module_mu}), the propagation 
equations~\refp{eq:field_Lmod3}, \refp{eq:field_Rmod3} depend only on the two signs 
$\varepsilon\equiv \varepsilon_{\lambda}\varepsilon_{\mu}$, $\varepsilon_{\gamma}$, 
and on the two real parameters $\module{\nu}$, $\theta$. 
The space-time units, $a$ and $\tau$, are again introduced so that $c_0=a/\tau$ 
is a microscopic velocity. The choices $\varepsilon=-1$ and $\varepsilon_{\gamma}=1$ leads 
to the identification with the Dirac equation in the Majorana-Weyl representation~\refp{eq:contdirac}

\begin{eqnarray}
\lefteqn{ \frac{1}{2\tau}\left(\psi_L(n,t+\tau) - \psi_{L}(n,t-\tau)\right) \: =} \nonumber \\
&&	-\left(\frac{2\sin\theta}{2+{\module{\nu}}^{2}}\right)\frac{a}{\tau}
	\left(\frac{\psi_{L}(n+a,t)-\psi_{L}(n-a,t)}{2a}\right) - 
	\frac{i}{\tau}\left(\frac{\sin\theta{\module{\nu}}^{2}}
	{2+{\module{\nu}}^{2}}\right) \, \psi_{R}(n,t), 
\label{eq:field_Lmod4} \\
&& \nonumber \\
\lefteqn{ \frac{1}{2\tau}\left(\psi_R(n,t+\tau) - \psi_{R}(n,t-\tau)\right) \: = } \nonumber \\
&&	-\left(\frac{2\sin\theta}{2+{\module{\nu}}^{2}}\right)\frac{a}{\tau}
	\left(\frac{\psi_{R}(n+a,t)-\psi_{R}(n-a,t)}{2a}\right)-
	\frac{i}{\tau}\left(\frac{\sin\theta{\module{\nu}}^{2}}
	{2+{\module{\nu}}^{2}}\right) \, \psi_{L}(n,t).
\label{eq:field_Rmod4}
\end{eqnarray}
Comparison between~\refp{eq:field_Lmod4}, \refp{eq:field_Rmod4} 
and~\refp{eq:contdirac} provides the velocity $c$ and the mass $m$ of the chiral fields

\begin{eqnarray}
c & = & c_0\frac{2\sin\theta}{2+{\module{\nu}}^{2}}, \label{eq:id_dirac_1} \\
\frac{mc^2}{\hbar} & = & \frac{1}{\tau}
	\frac{\sin\theta{\module{\nu}}^{2}}{2+{\module{\nu}}^{2}}.
\label{eq:id_dirac_2}
\end{eqnarray}

\subsubsection*{Exact form of the scattering matrices}
The $L,R$ matrix coefficients~\refp{def:chiral_scattmod1}, \refp{def:chiral_scattmod2} depend
only on $\mu$, on the ratios $\mu_{1,L}/\mu=\mu_{2,L}/\mu=\varepsilon_{\mu}$, 
$\lambda_{2,L}/\lambda_{1,L}=i\varepsilon_{\lambda}=-i\varepsilon_{\mu}$, 
and $\nu$. However the phase of $\nu$ is fixed
because $\nu^{\ast}/\nu=i$, thanks to the third equation 
of~\refp{eq:is_matrixcoef1} and to~\refp{eq:is_matrixcoef2}.
Therefore, the $L,R$ scattering matrices are parametrized by three
real parameters~: $\varepsilon_{\mu}, \theta$ and $\module{\nu}$.
We introduce a complex number $\alpha$ defined by 

\begin{equation}
\alpha^2=i\varepsilon_{\mu}, \label{eq:alpha}
\end{equation}
so that the scattering matrices read

\begin{equation}
\begin{array}{lclcl}
S_L & = & -\displaystyle{\frac{2\sin\theta}{2+{\module{\nu}}^{2}}}\alpha^2\left(
\begin{array}{ccc}
1 & -\alpha^2 & \nu \\
\\
\alpha^2 & 1 & \alpha^2\nu \\
\\
\alpha^2\nu & \nu & \alpha^2{\nu}^{2}
\end{array}
\right) & + & \mu \, \mbox{diag}\left(\varepsilon_{\mu},\varepsilon_{\mu},1\right), 
\end{array}
\label{eq:matrice_chiralL}
\end{equation}

\begin{equation}
\begin{array}{lclcl}
S_R & = & -\displaystyle{\frac{2\sin\theta}{2+{\module{\nu}}^{2}}}\alpha^2\left(
\begin{array}{ccc}
1 & \alpha^2 & \alpha^2\nu \\ 
\\
-\alpha^2 & 1 & \nu \\
\\
\nu & \alpha^2\nu & \alpha^2{\nu}^{2}
\end{array}
\right) & + & \mu \, \mbox{diag}\left(\varepsilon_{\mu},\varepsilon_{\mu},1\right). 
\end{array}
\label{eq:matrice_chiralR}
\end{equation} 
Lastly, one finds easily the following relations

\begin{equation}
\frac{\lambda_{1,L}^{\ast}}{\lambda_{1,L}}=-i\mu, \esp 
\frac{\lambda_{2,L}^{\ast}}{\lambda_{2,L}}=i\mu,
\label{eq:phase_lambda}
\end{equation} 
and

\begin{equation}
\lambda_{1,R}=-i\varepsilon_{\mu}\lambda_{1,L}, \esp \lambda_{2,R}=i\varepsilon_{\mu}\lambda_{2,L},
\esp \lambda_{3,R}=\lambda_{3,L},
\label{eq:link_lambdaLR}
\end{equation}
which will be used later in the standard representation (section~\ref{symet_sta}).

%*****************************************************************************

\section{Standard representation}~\label{sta}
In this section, the ideas developed in the construction of the current model
for the Majorana-Weyl representation are reproduced for the standard 
representation in $1+1$ space-time dimensions. Let us recall the Hamiltonian form 
of the Dirac equation in the standard representation 

\begin{equation}
i\frac{\partial \Psi}{\partial t} = \left[-ic \sigma_1 
\frac{\partial}{\partial x} + \frac{m c^2}{\hbar} \sigma_3 \right] \Psi,
\label{eq:contdirac_sta}
\end{equation}
where $\Psi=(\psi_+ \:\: \psi_-)^{T}$ is the two component spinor field
in the standard representation. Surprisingly, it turns out that the scattering 
matrices cannot be computed by a direct construction as in the Majorana-Weyl representation. 
More precisely, it appears that the natural symmetries of the Dirac equation in the 
standard representation transposed to the scattering process lead to contradictions 
in the computation of the scattering matrices. However, if we take advantage of 
its linear connection with the Majorana-Weyl representation, we are 
able to build indirectly the current model for the standard representation. 

\subsection{Basic definitions~: currents and fields}~\label{dyna_sta}
\begin{figure}[hbt]
\centerline{\epsffile{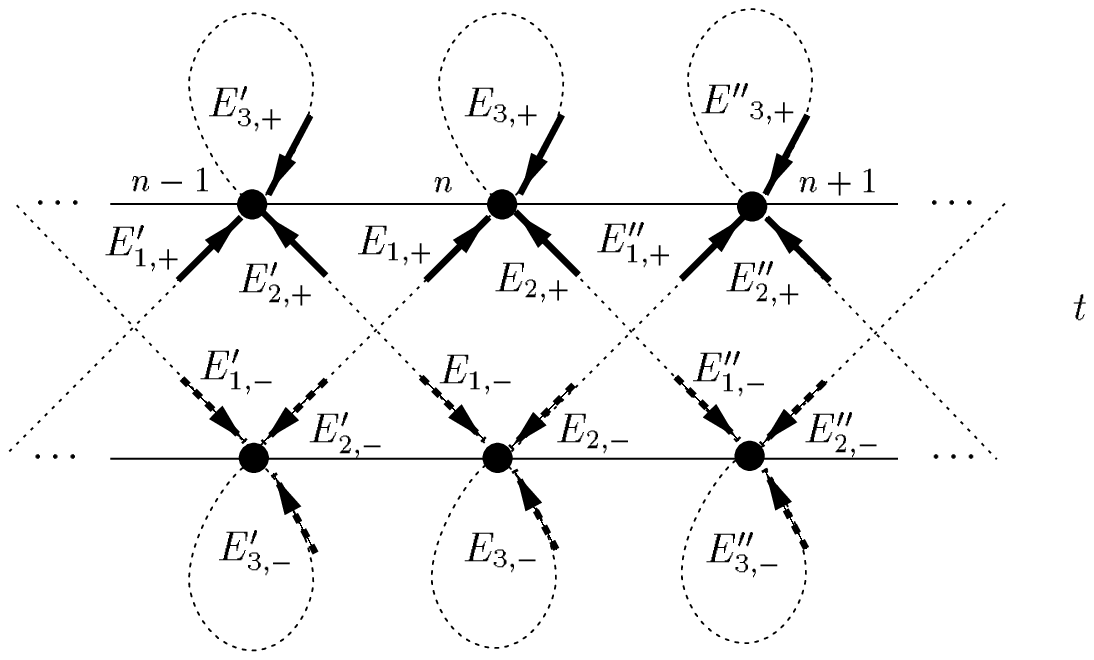}}
\caption{Sketch of the chain in the standard representation. The upper and lower subchains carry
the $+$ and $-$ currents respectively.}
\label{fig:modele_sta}
\end{figure}

The two current models constructed for the standard
and the Majorana-Weyl representations share many features.
Indeed, two kinds of incident and outgoing currents are defined on the bonds and on the nodes
of a regular chain as the goal is to describe the propagation equations obeyed by the
two scalar fields $\psi_{+}$ and $\psi_{-}$. The lattice mesh size $a$ and the elementary 
time step $\tau$ are taken equal to unity. Transposing the notations of 
section~\ref{dyna_maj}, incident currents are denoted by $E_{k,+}$, $E_{l,-}$, 
whereas outgoing currents are denoted by $S_{k',+}$, $S_{l',-}$. Three incident 
currents of each kind live on each node of the chain depicted in 
Figure \ref{fig:modele_sta}. $E_{1,2,\pm}$ denotes the $\pm$ bond-currents and 
$E_{3,\pm}$ denotes the node-currents. The chain schematically represented
in Figure \ref{fig:modele_sta} is deduced from the chain depicted 
in Figure \ref{fig:modele_maj1}, for the Majorana-Weyl representation, by replacing the 
$L$ currents by $+$ currents and the $R$ currents by $-$ currents. 
   
The dynamics of the $\pm$ currents is similar to the dynamics defined for the 
$L,R$ currents in the Majorana-Weyl representation. The $\pm$ currents obey
a discrete Huygens principle on the chain which decomposes in two steps~: 
a propagation step and a scattering step. At a given time $t$, all the $\pm$ currents 
undergo the dynamics simultaneously. The incident currents $E_{j,+}$, $E_{l,-}$, 
are scattered instantaneously on each node of the chain to become outgoing currents
$S_{i,+}$, $S_{k,-}$, which are linear superposition of those incident currents

\begin{eqnarray}
S_{i,+}(n,t) & = & \sum_{j=1}^{3} s_{ij,+} E_{j,+}(n,t), 
\esp i=1,2,3,\label{eq:scatt_sta+}\\
S_{k,-}(n,t) & = & \sum_{l=1}^{3} s_{kl,-} E_{l,-}(n,t),
\esp k=1,2,3.\label{eq:scatt_sta-}
\end{eqnarray}
where $s_{ij,+}$, $s_{kl,-}$ are the complex elements of the scattering matrices $S_{\pm}$
attached to each node of the chain. Then, in a unit time step, the $\pm$ outgoing currents
propagate from one node to the nearest-neighbor ones to become incident currents at 
time $t+1$. Using the notations previously introduced in Figure \ref{fig:def_bond},
the propagation rules read

\begin{equation}
\left\{
\begin{array}{lccccl}
E_{k,+}(n,t+1) = S_{\overline{k},-}(n_k,t), & k=1,2, &&&& 
E_{3,+}(n,t+1) = S_{3,+}(n,t), \\
\\
E_{l,-}(n,t+1) = S_{\overline{l},+}(n_l,t), & l=1,2, &&&&
E_{3,-}(n,t+1) = -S_{3,-}(n,t).
\end{array}
\right.
\label{eq:proprule_sta}
\end{equation} 

\begin{figure}[hbt]
\centerline{\epsffile{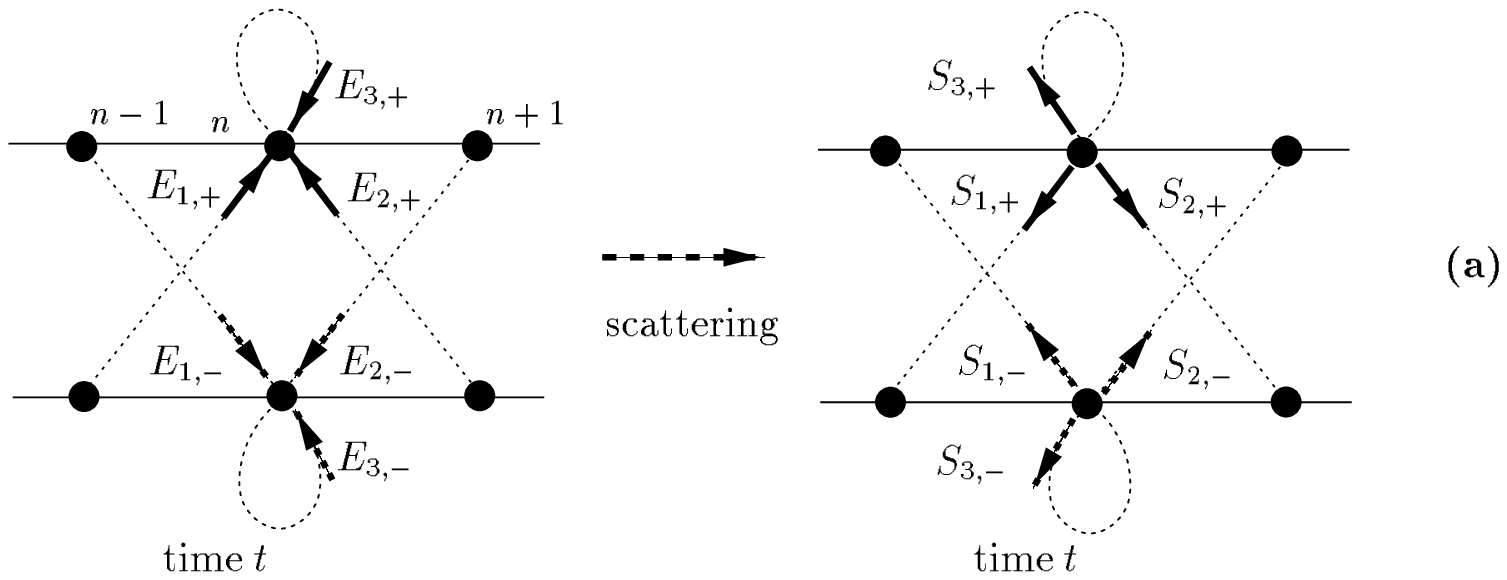}}
\centerline{\epsffile{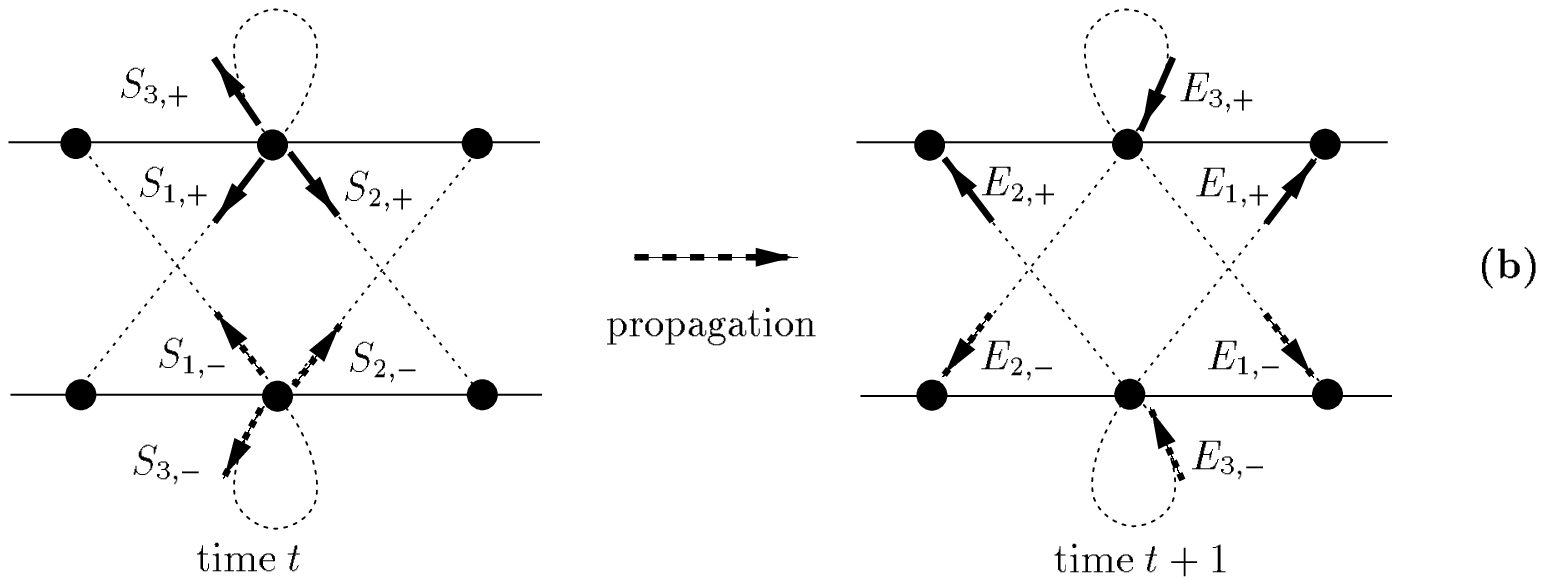}}
\caption{Scattering step {\bf (a)} and propagation step 
{\bf (b)} for the $+$ and $-$ currents.}
\label{fig:difprop_sta}
\end{figure}

Contrary to the construction of the current-model in the Majorana-Weyl 
representation (Fig.\ \ref{fig:difprop_chiral}) the transmutation 
affects solely the bond-currents and not the node-currents. The choice
of such propagation rules stems from the structure of the Dirac equation~\refp{eq:contdirac_sta}
in which the spatial derivative $\sigma_{1}\partial/\partial x$ couples $\psi_{+}$ and $\psi_{-}$
while the last term $(mc^2/\hbar)\sigma_{3}$ does not. Additionally, a minus sign 
affects the propagation of the $-$ node-current. The reason for introducing this minus
sign will be clarified at the end of the construction (see section~\ref{eq_sta}). 
The dynamics of the $\pm$ currents is summarized in Figure \ref{fig:difprop_sta}.

Finally, similarly to the definitions \refp{def:chiral_comp_a}, \refp{def:chiral_comp_b} 
of the chiral fields in the Majorana-Weyl representation,
we suppose that the scalar fields $\psi_+$ and $\psi_-$ 
are defined on each site of the chain as complex linear superpositions of 
$+$ or $-$ incident currents respectively

\begin{eqnarray}
\psi_+(n,t) & = & \sum_{k=1}^{3} \lambda_{k,+} \, E_{k,+}(n,t), 
\label{def:standard_comp_a}\\
\psi_-(n,t) & = & \sum_{l=1}^{3} \lambda_{l,-} \, E_{l,-}(n,t),
\label{def:standard_comp_b}
\end{eqnarray}
where $\lambda_{k,+}$, $\lambda_{l,-}$ are complex numbers. 

\subsection{Discrete propagation 
equations linking $\psi_+$ and $\psi_-$}~\label{eq_sta}
Given the local rules which govern the dynamics of the $\pm$ currents, our aim is to 
derive two closed propagation equations involving the fields $\psi_{\pm}$

\begin{eqnarray}
\psi_+(n,t+1) & = & f_+\left( \psi_+(n',t'),\psi_-(n',t'),
\psi_+(n'',t''),\psi_-(n'',t''),\ldots \right),
\label{eq:wave_field_+} \\
\psi_-(n,t+1) & = & f_-\left( \psi_-(n',t'),\psi_+(n',t'),
\psi_-(n'',t''),\psi_+(n'',t''),\ldots \right),
\label{eq:wave_field_-}
\end{eqnarray}
where the fields $\psi_{\pm}$ on node $n$ and at time $t+1$ are functions
of both fields on the same node and/or on neighboring nodes $n'$, $n''$, $\ldots$, at 
previous times $t'$, $t''$, $\ldots$ The necessary conditions constraining the scattering 
matrices $S_{\pm}$ for the equations~\refp{eq:wave_field_+}, \refp{eq:wave_field_-}
to be closed remain the same as in the Majorana-Weyl construction (see 
section~\ref{eq_maj}). So that the scattering matrices are of the form  

\begin{eqnarray}
S_{+} & = & P_{+} +\mbox{diag}(\mu_{1,+},\mu_{2,+},\mu_{3,+}),
\label{def:standard_scatt_a} \\
S_{-} & = & P_{-} +\mbox{diag}(\mu_{1,-},\mu_{2,-},\mu_{3,-}),
\label{def:standard_scatt_b}
\end{eqnarray}
where $P_{\pm}$ are matrices which are generically given,
for each kind of current, by \refp{def:chiral_scatt_c} (with $\alpha=\pm$). The above form of
the scattering matrices $S_{+}$ and $S_{-}$ implies for the outgoing currents
the following equations

\begin{eqnarray}
S_{k,+} & = & \rho_{k,+} \psi_{+} + \mu_{k,+} E_{k,+},  \label{eq:outcurrent+} \\
S_{k,-} & = & \rho_{k,-} \psi_{-} + \mu_{k,-}E_{k,-}, \esp k=1,2,3. \label{eq:outcurrent-} 
\end{eqnarray}
The rest of the derivation is almost the same as for the propagation 
equations~\refp{eq:field_L}, \refp{eq:field_R} in the Majorana-Weyl representation.

Using the propagation rules~\refp{eq:proprule_sta} and 
the definition~\refp{def:standard_comp_a}, the field $\psi_{+}$ on node $n$ 
at time $t+1$ can be expressed in terms of $\pm$ outgoing currents, 
defined at the previous time $t$ and on the neighbor nodes $n_{k}$,

\begin{equation}
\psi_+(n,t+1) = \sum_{k=1,2} \lambda_{k,+} 
\, S_{\overline{k},-}(n_k,t) + \lambda_{3,+} \, S_{3,+}(n,t). 
\label{eq:psi+_int_1}
\end{equation} 
Equation~\refp{eq:psi+_int_1} is sketched below

\begin{figure}[!hbt]
\begin{center}
\epsffile{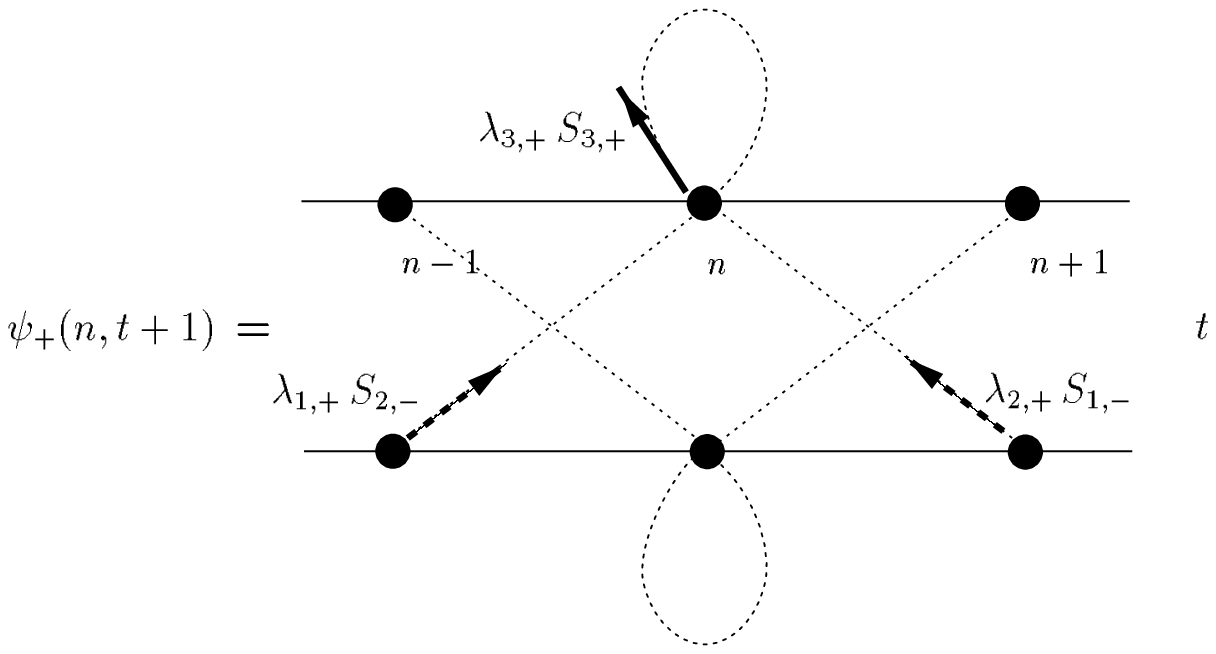}
\end{center}
\end{figure}
The outgoing currents entering equation~\refp{eq:psi+_int_1} were
scattered instantaneously at time $t$~: $S_{1,-}$, $S_{2,-}$,
and $S_{3,+}$, result from linear combinations of ingoing currents
of the form~\refp{eq:outcurrent+} and \refp{eq:outcurrent-}. As a consequence, 
$\psi_+(n,t+1)$ is represented schematically in terms of $\pm$ ingoing 
currents at time $t$ as depicted as follows~:

\begin{figure}[!hbt]
\epsffile{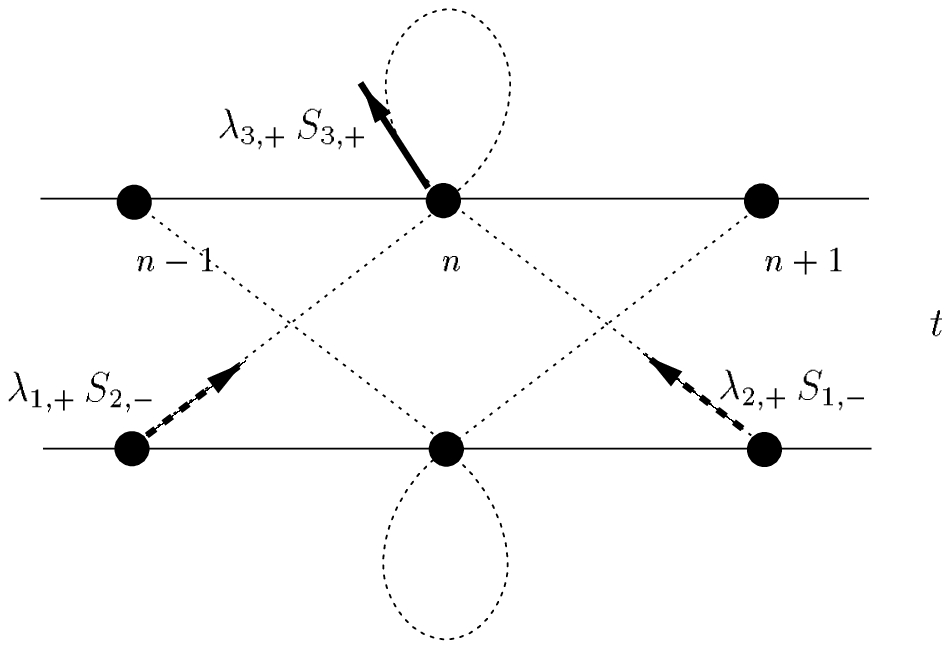}
\end{figure}
\newpage
\begin{figure}[!hbt]
\begin{center}
\epsffile{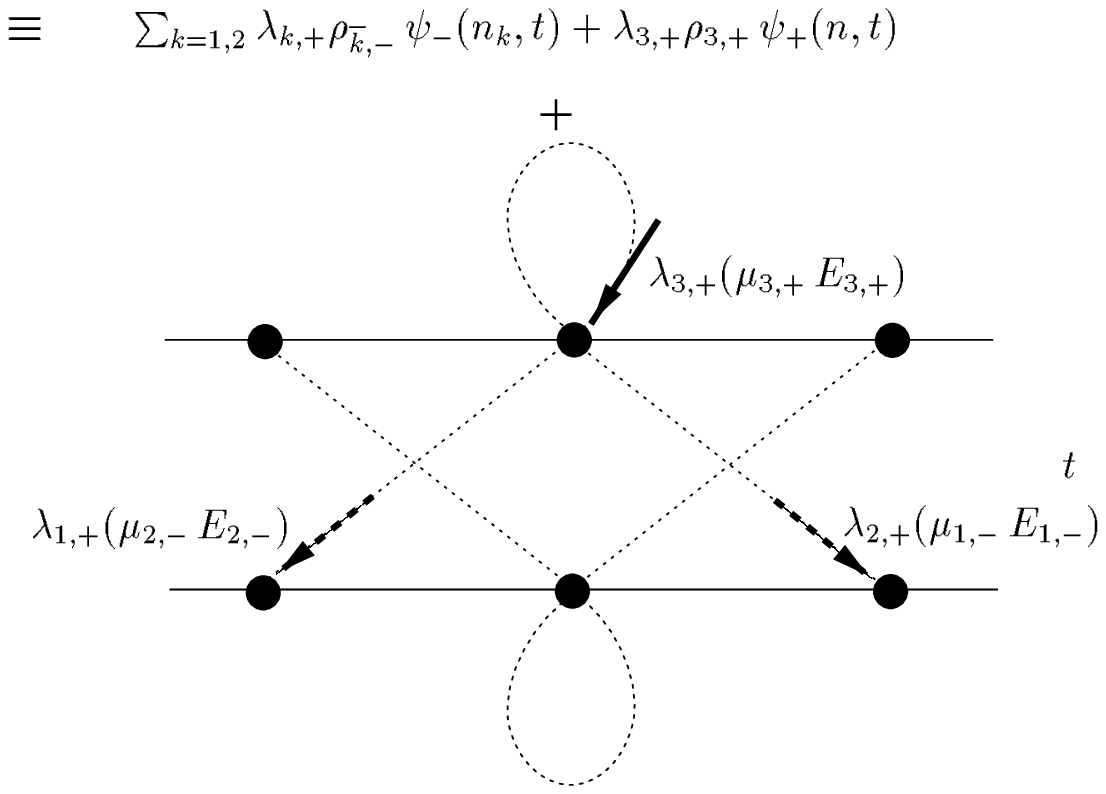}
\end{center}
\end{figure}
which reads analytically

\begin{eqnarray}
\lefteqn{\psi_+(n,t+1) = \sum_{k=1,2} \lambda_{k,+} \rho_{\overline{k},-} 
\, \psi_-(n_k,t) + \lambda_{3,+} \rho_{3,+} \, \psi_+(n,t)} \nonumber \\
& & +\underbrace{\sum_{k=1,2}\lambda_{k,+}\mu_{\overline{k},-}    
\, E_{\overline{k},-}(n_k,t) + \lambda_{3,+} \mu_{3,+} 
\, E_{3,+}(n,t).}_{\mbox{current term}} \label{eq:psi+_int_2}
\end{eqnarray}
In virtue of the propagation and transmutation rules, the current term of 
equation~\refp{eq:psi+_int_2} is deduced from three $\pm$ outgoing
currents that were defined on node $n$ and at the previous time $t-1$

\begin{equation}
\mbox{current term} = \sum_{k=1,2}\lambda_{k,+}\mu_{\overline{k},-}\, S_{k,+}(n,t-1)
+\lambda_{3,+}\mu_{3,+}\, S_{3,+}(n,t-1).
\label{eq:psi+_int_3}
\end{equation}
Applying the scattering rules to the outgoing currents entering
the last equation gives rise to a field $\psi_{+}$ and to three $+$ 
incident currents, at time $t-1$

\begin{eqnarray}
\lefteqn{\mbox{current term} = 
	\left(\sum_{k=1,2} \lambda_{k,+} \mu_{\overline{k},-} \rho_{k,+} 
        +\lambda_{3,+} \mu_{3,+} \rho_{3,+}\right) \, \psi_{+}(n,t-1) } \nonumber \\ 
\nonumber \\
& &     +\sum_{k=1,2} \lambda_{k,+} \mu_{\overline{k},-} \mu_{k,+} 
	\, E_{k,+}(n,t-1) 
	+\lambda_{3,+} \mu_{3,+}^2 \, E_{3,+}(n,t-1). \label{eq:psi+_int_4}
\end{eqnarray}
At this stage, we see that in order to close equation~\refp{eq:psi+_int_2} 
at time $t-1$, it is sufficient to impose

\begin{equation}
\left\{
\begin{array}{ccl}
\mu_{k,+} \mu_{\overline{k},-} & = & \mu_{+}^2, \esp k=1,2, \\
\\
\mu_{3,+}^2  & = & \mu_{+}^2, 
\end{array}
\right.
\label{eq:condsuf+}
\end{equation}
where we have introduced the constant $\mu_{+}^2$. We end up the derivation of the closed 
propagation equation satisfied by $\psi_+$
by plugging the current term~\refp{eq:psi+_int_4} into equation~\refp{eq:psi+_int_2}, 
supplied with the two conditions~\refp{eq:condsuf+},

\begin{eqnarray}
\lefteqn{\psi_+(n,t+1) - \mu_{+}^2\left( 1+\sum_{k=1}^{3} \frac{\lambda_{k,+} 
\rho_{k,+}}{\mu_{k,+}}\right) \, \psi_{+}(n,t-1) =} \nonumber \\
& &      \sum_{k=1,2} \lambda_{k,+} \rho_{\overline{k},-} \, \psi_{-}(n_k,t) 
        + \lambda_{3,+} \rho_{3,+} \, \psi_{+}(n,t).
\label{eq:field_+}
\end{eqnarray}
The propagation equation for $\psi_{-}$ can be directly deduced 
from equation~\refp{eq:field_+} by inverting all the subscripts $+$ and $-$, except
that we must take care of the minus sign involved in the
propagation step~\refp{eq:proprule_sta} of the $-$ node current. 
The propagation equation for $\psi_{-}$ is obtained in its final form 
by noting that the minus sign appears first in the expression of
$\psi_{-}(n,t+1)$ involving $\pm$ outgoing currents (equation analogous to 
equation~\refp{eq:psi+_int_1})

\[
\psi_-(n,t+1) = \sum_{k=1,2} \lambda_{k,-} 
\, S_{\overline{k},+}(n_k,t) - \lambda_{3,-} \, S_{3,-}(n,t). 
\]
Then, according to the scattering step, the last term of this equation
gives rise to the term~: $-\lambda_{3,-}\rho_{3,-} \, \psi_{-}(n,t)$,
which is analogous to the term $\lambda_{3,+}\rho_{3,+} \, \psi_{+}(n,t)$ in equation
\refp{eq:field_+}. Finally, one can check that this term is the only one affected
by the minus sign so that the propagation equation for $\psi_{-}$ reads  

\begin{eqnarray}
\lefteqn{\psi_-(n,t+1) - \mu_{-}^2\left( 1+\sum_{k=1}^{3} \frac{\lambda_{k,-} 
\rho_{k,-}}{\mu_{k,-}}\right) \, \psi_{-}(n,t-1) =} \nonumber \\
& &      \sum_{k=1,2} \lambda_{k,-} \rho_{\overline{k},+} \, \psi_{+}(n_k,t) 
        - \lambda_{3,-} \rho_{3,-} \, \psi_{-}(n,t). 
\label{eq:field_-}
\end{eqnarray}
Equation~\refp{eq:field_-} is supplied with the following sufficient conditions

\begin{equation}
\left\{
\begin{array}{ccl}
\mu_{k,-} \mu_{\overline{k},+} & = & \mu_{-}^2, \esp k=1,2, \\
\\
\mu_{3,-}^2  & = & \mu_{-}^2.
\end{array} 
\right.
\label{eq:condsuf-}
\end{equation}
Comparison between~\refp{eq:condsuf+} and \refp{eq:condsuf-} shows that $\mu_{+}^2=\mu_{-}^2$ which
will be noted $\mu^2$ in the following. 

\subsection{Tranformation equations from the Majorana-Weyl to the standard 
representation}~\label{symet_sta}
Implementing symmetries in the current model enables one 
to determine uniquely the form of the scattering matrices and 
of the propagation equations. Additionally, it provides a stability 
condition on the dynamics of the currents, which is fulfilled 
if the scattering matrices are unitary. Although this method was applied 
successfully in order to build the current model for the Majorana-Weyl 
construction (sections~\ref{symet_maj} and~\ref{scatt_maj}), it turns out that
it fails for the standard representation. We thus proceed by using an indirect 
method in obtaining the $\pm$ scattering matrices that relies on the linear 
transformation between the Majorana-Weyl and the standard representations. 
Contrary to the direct method, we show that there exist several possible expressions
for the scattering matrices. The linear correspondence between the Majorana-Weyl and the 
standard representation reads

\begin{equation}
\left\{
\begin{array}{ccl}
\psi_{+} & = & \psi_{L}+\psi_{R}, \\
\psi_{-} & = & \psi_{L}-\psi_{R}.
\end{array}
\right.
\label{eq:connection}
\end{equation}
According to the definitions of the fields $\psi_{L}$ and $\psi_{R}$
(Eqs.\ \refp{def:chiral_comp_a}, \refp{def:chiral_comp_b}), $\psi_+$ and $\psi_-$ 
in~\refp{eq:connection} become

\begin{eqnarray}
\psi_{+} & = & \sum_{k=1}^{3} \lambda_{k,L} E_{k,L} + \sum_{k=1}^{3} \lambda_{k,R} E_{k,R}, 
\label{eq:detail+} \\
\psi_{-} & = & \sum_{k=1}^{3} \lambda_{k,L} E_{k,L} - \sum_{k=1}^{3} \lambda_{k,R} E_{k,R}.
\label{eq:detail-}
\end{eqnarray}
If we remember that the $\lambda_{k,L}$ and the $\lambda_{k,R}$ are linked together
by the relations~\refp{eq:link_lambdaLR}, then it is possible to express $\psi_{+}$ and $\psi_{-}$
as a function of the $\lambda_{k,L}$ or of the $\lambda_{k,R}$ uniquely. Let us first
put the emphasis on the $L$ parameters rather than on the $R$ ones. 
For this purpose, let us write~\refp{eq:detail+}, \refp{eq:detail-} in the following form

\begin{eqnarray}
\lefteqn{\psi_{+} = \frac{1}{2}\sum_{k=1,2}\left[
(\lambda_{k,L}-i\varepsilon_{\mu}\lambda_{k,R})(E_{k,L}+i\varepsilon_{\mu}E_{k,R})
+(\lambda_{k,L}+i\varepsilon_{\mu}\lambda_{k,R})(E_{k,L}-i\varepsilon_{\mu}E_{k,R})\right] } \nonumber \\
&&	+\frac{1}{2}\left[(\lambda_{3,L}+\lambda_{3,R})(E_{3,L}+E_{3,R})+
	(\lambda_{3,L}-\lambda_{3,R})(E_{3,L}-E_{3,R})\right], 
\label{eq:connect_field+_1}\\
\lefteqn{\psi_{-} = \frac{1}{2}\sum_{k=1,2}\left[
(\lambda_{k,L}+i\varepsilon_{\mu}\lambda_{k,R})(E_{k,L}+i\varepsilon_{\mu}E_{k,R})
+(\lambda_{k,L}-i\varepsilon_{\mu}\lambda_{k,R})(E_{k,L}-i\varepsilon_{\mu}E_{k,R})\right] } \nonumber \\
&&	+\frac{1}{2}\left[(\lambda_{3,L}-\lambda_{3,R})(E_{3,L}+E_{3,R})+
	(\lambda_{3,L}+\lambda_{3,R})(E_{3,L}-E_{3,R})\right].
\label{eq:connect_field-_1}
\end{eqnarray}
If we plug~\refp{eq:link_lambdaLR} into the above equations, we see that terms are cancelled
in each equation so that $\psi_{+}$ and $\psi_{-}$ become

\begin{eqnarray*}
\psi_{+} & = & \lambda_{1,L}(E_{1,L}-i\varepsilon_{\mu}E_{1,R})+
\lambda_{2,L}(E_{2,L}+i\varepsilon_{\mu}E_{2,R})+\lambda_{3,L}(E_{3,L}+E_{3,R}), \\
\psi_{-} & = & \lambda_{1,L}(E_{1,L}+i\varepsilon_{\mu}E_{1,R})+
\lambda_{2,L}(E_{2,L}-i\varepsilon_{\mu}E_{2,R})+\lambda_{3,L}(E_{3,L}-E_{3,R}).
\end{eqnarray*}
Now, identifying $\lambda_{k,+}$ and $\lambda_{k,-}$ with $\lambda_{k,L}$, $k=1,2,3$, 
imposes the following relations between the $\pm$ incident currents and the $L,R$ ones

\begin{equation}
\left\{
\begin{array}{ccl}
E_{1,\pm} & = & E_{1,L}\mp i\varepsilon_{\mu}E_{1,R}, \\
E_{2,\pm} & = & E_{2,L}\pm i\varepsilon_{\mu}E_{2,R}, \\
E_{3,\pm} & = & E_{3,L}\pm E_{3,R}.
\end{array}
\right.
\label{eq:connect_L}
\end{equation}
Assuming that the same relationships hold between the $\pm$ and $L,R$ outgoing currents,
one finds that by substituting the propagation rules~\refp{eq:proprule_chiral1}, \refp{eq:proprule_chiral2} 
for the $L,R$ currents in \refp{eq:connect_L}, the propagation 
rules~\refp{eq:proprule_sta} for the $\pm$ currents are recovered. The scattering matrices 
$S_{\pm}$ are also deduced from \refp{eq:connect_L}, by using the 
expressions of the scattering matrices in the Majorana-Weyl representation~\refp{eq:matrice_chiralL}, 
\refp{eq:matrice_chiralR} leading to
 
\begin{equation}
S_{\pm} = S_{L}.
\label{eq:connect_scattL}
\end{equation}
If instead, we put the emphasis on the $R$ currents rather than on the $L$ ones, the $\pm$ fields
are obtained through the equations~\refp{eq:connect_field+_1}, \refp{eq:connect_field-_1}  
where the roles of the $L$ and $R$ subscripts are exchanged. If we plug~\refp{eq:link_lambdaLR}
in those two equations, $\psi_{\pm}$ becomes

\begin{eqnarray*}
\psi_{+} & = & \lambda_{1,R}(E_{1,R}+i\varepsilon_{\mu}E_{1,L})+
\lambda_{2,R}(E_{2,R}-i\varepsilon_{\mu}E_{2,L})+\lambda_{3,R}(E_{3,R}+E_{3,L}), \\
\psi_{-} & = & -\lambda_{1,R}(E_{1,R}-i\varepsilon_{\mu}E_{1,L})-
\lambda_{2,R}(E_{2,R}+i\varepsilon_{\mu}E_{2,L})-\lambda_{3,R}(E_{3,R}-E_{3,L}).
\end{eqnarray*} 
Then, by choosing $\lambda_{k,+}=-\lambda_{k,-}=\lambda_{k,R}$, $k=1,2,3$, we fix the  
relations between the $\pm$ incident currents and the $L,R$ ones

\begin{equation}
\left\{
\begin{array}{ccl}
E_{1,\pm} & = & E_{1,R} \pm i\varepsilon_{\mu}E_{1,L}, \\
E_{2,\pm} & = & E_{2,R}\mp i\varepsilon_{\mu}E_{2,L}, \\
E_{3,\pm} & = & E_{3,R}\pm E_{3,L}.
\end{array}
\right.
\label{eq:connect_R}
\end{equation}
Again, if we assume that the same relationships hold between the outgoing currents of both 
representations, the propagation rules of the standard representation are recovered and 
the $\pm$ scattering matrices read

\begin{equation}
S_{\pm} = S_{R}.
\label{eq:connect_scattR}
\end{equation}
Finally, we present a more symmetric transformation between the two representations
where the $L$ and $R$ currents are treated on the same footing.
The starting point is to write the $\pm$ fields~\refp{eq:detail+}, \refp{eq:detail-} in 
the following form

\begin{eqnarray}
\psi_{+} & = & \alpha^{\ast}\lambda_{1,L}(\alpha E_{1,L}+\alpha^{\ast} E_{1,R})
	+\alpha\lambda_{2,L}(\alpha^{\ast} E_{2,L}+\alpha E_{2,R})
	+\lambda_{3,L}(E_{3,L}+E_{3,R}), \label{eq:connect_field+_2} \\
\psi_{-} & = & \alpha\lambda_{1,L}(\alpha^{\ast} E_{1,L}+\alpha E_{1,R})
	+\alpha^{\ast}\lambda_{2,L}(\alpha E_{2,L}+\alpha^{\ast} E_{2,R})
	+\alpha^2\lambda_{3,L}({\alpha^{\ast}}^{2}E_{3,L}+\alpha^{2}E_{3,R}),
\label{eq:connect_field-_2}
\end{eqnarray} 
where $\alpha$ is defined by $\alpha^2=i\varepsilon_{\mu}$ (Eq.\ \refp{eq:alpha}). 
The identities~:
$\alpha^{\ast}\lambda_{1,L}=\alpha\lambda_{1,R}$, 
$\alpha\lambda_{1,L}=-\alpha^{\ast}\lambda_{1,R}$, 
$\alpha^{\ast}\lambda_{2,L}=-\alpha\lambda_{2,R}$,
$\alpha\lambda_{2,L}=\alpha^{\ast}\lambda_{2,R}$, which follow from~\refp{eq:link_lambdaLR},
insure that neither the $L$ incident currents nor the $R$ incident currents are preferred in 
equations~\refp{eq:connect_field+_2}, \refp{eq:connect_field-_2}. 

The choice

\begin{equation}
\left\{
\begin{array}{cclccclcccl}
\lambda_{1,+} & = & \alpha^{\ast}\lambda_{1,L}, && \lambda_{2,+} & = & \alpha\lambda_{2,L}, && 
\lambda_{3,+} & = & \lambda_{3,L}, \\
\lambda_{1,-} & = & \alpha\lambda_{1,L}, && \lambda_{2,-} & = & \alpha^{\ast}\lambda_{2,L}, &&
\lambda_{3,-} & = & \alpha^2\lambda_{3,L}, 
\end{array}
\right.
\label{eq:lambda_sym}
\end{equation}
fixes $\rho_{+}$ and $\rho_{-}$

\begin{equation}
\left\{
\begin{array}{cclccclcccl}
\rho_{1,+} & = & \alpha\rho_{1,L}, && \rho_{2,+} & = & \alpha^{\ast}\rho_{2,L}, &&
\rho_{3,+} & = & \rho_{3,L}, \\
\rho_{1,-} & = & \alpha^{\ast}\rho_{1,L}, && \rho_{2,-} & = & \alpha\rho_{2,L}, &&
\rho_{3,-} & = & {\alpha^{\ast}}^2\rho_{3,L}, 
\end{array}
\right.
\label{eq:rho_sym}
\end{equation}
and the relations between the $\pm$ incident currents and the $L,R$ ones

\begin{equation}
\left\{
\begin{array}{cclcccl}
E_{1,+} & = & \alpha E_{1,L}+\alpha^{\ast} E_{1,R}, && E_{1,-} & = & 
\alpha^{\ast} E_{1,L}+\alpha E_{1,R}, \\
E_{2,+} & = & \alpha^{\ast} E_{2,L}+\alpha E_{2,R}, && E_{2,-} & = & 
\alpha E_{2,L}+\alpha^{\ast} E_{2,R}, \\
E_{3,+} & = & E_{3,L}+E_{3,R}, && E_{3,-} & = & {\alpha^{2}}^{\ast}E_{3,L}+\alpha^{2}E_{3,R}.
\end{array}
\right.
\label{eq:connect_mix}
\end{equation}
Assuming that the transformation rules~\refp{eq:connect_mix} link also the outgoing currents 
of each representation, we recover the propagation rules of the standard representation and
derive the $\pm$ scattering matrices 

\begin{equation}
\begin{array}{lclcl}
S_+ & = & \displaystyle{-\frac{2\sin\theta}{2+{\module{\nu}}^{2}}}\alpha^{2}\left(
\begin{array}{ccc}
1 & 1 & \alpha\nu \\
\\
1 & 1 & \alpha\nu \\
\\
\alpha\nu & \alpha\nu & \alpha^2{\nu}^{2}
\end{array}
\right) & + & \mu \, \mbox{diag}\left(\varepsilon_{\mu},\varepsilon_{\mu},1\right),
\end{array}
\label{eq:matrice_mix+}
\end{equation}

\begin{equation}
\begin{array}{lclcl}
S_- & = & \displaystyle{-\frac{2\sin\theta}{2+{\module{\nu}}^{2}}}\alpha^{2}\left(
\begin{array}{ccc}
1 & -1 & \alpha\nu \\
\\
-1 & 1 & -\alpha\nu \\
\\
\alpha\nu & -\alpha\nu & \alpha^2{\nu}^{2}
\end{array}
\right) & + & \mu \, \mbox{diag}\left(\varepsilon_{\mu},\varepsilon_{\mu},1\right).
\end{array}
\label{eq:matrice_mix-}
\end{equation}
Within this symmetric transformation both scattering matrices are different
from $S_L$ and $S_R$. Meanwhile, each one satisfies the symmetries~: 
$S_{\pm}={}^TS_{\pm}$, $S_{\pm}^{-1}=S_{\pm}^{\ast}$, and are indeed unitary matrices. 
The first property, namely $S_{\pm}={}^TS_{\pm}$, can be related to a reciprocity
invariance of the scattering process, whether the second property, 
$S_{\pm}^{-1}=S_{\pm}^{\ast}$, can be related to a time reversal invariance of the scattering 
process. But we stress that neither this reciprocity property nor this time reversal symmetry are 
symmetries obeyed by the components $\psi_+$ and $\psi_-$ of the spinor field in the 
standard representation. Hence, contrary to the Majorana-Weyl representation, 
the symmetries of the current scattering process are different from the 
symmetries underlying the Dirac equation in the standard representation. This explains 
why we did not succeed in the direct construction of the standard representation.

Finally, using~\refp{eq:lambda_sym}, \refp{eq:rho_sym}, equations~\refp{eq:field_+}, 
\refp{eq:field_-} become 

\begin{eqnarray}
\lefteqn{ \frac{1}{2\tau}\left(\psi_+(n,t+\tau) - \psi_{+}(n,t-\tau)\right) \: =} \nonumber \\
&&      \left(\frac{2\sin\theta}{2+{\module{\nu}}^{2}}\right)\frac{a}{\tau}
        \left(\frac{\psi_{-}(n+a,t)-\psi_{-}(n-a,t)}{2a}\right)
        -\frac{i}{\tau}\left(\frac{\sin\theta{\module{\nu}}^{2}}
        {2+{\module{\nu}}^{2}}\right) \, \psi_{+}(n,t),
\label{eq:field_+end} \\
&& \nonumber \\
\lefteqn{ \frac{1}{2\tau}\left(\psi_-(n,t+\tau) - \psi_{-}(n,t-\tau)\right) \: = } \nonumber \\
&&      \left(\frac{2\sin\theta}{2+{\module{\nu}}^{2}}\right)\frac{a}{\tau}
        \left(\frac{\psi_{+}(n+a,t)-\psi_{+}(n-a,t)}{2a}\right)
        +\frac{i}{\tau}\left(\frac{\sin\theta{\module{\nu}}^{2}}
        {2+{\module{\nu}}^{2}}\right) \, \psi_{-}(n,t),
\label{eq:field_-end}
\end{eqnarray}
where

\begin{equation}
\left\{
\begin{array}{ccccl}
\mu^2 \left( 1 + \displaystyle{\sum_{k=1}^{3} \frac{\lambda_{k,+}\rho_{k,+}}{\mu_{k,+}}}
\right) & = & \mu^2 \left( 1 + \displaystyle{\sum_{k=1}^{3} 
\frac{\lambda_{k,-} \rho_{k,-}}{\mu_{k,-}}} \right) & = & 1, \\
\\
\lambda_{2,+} \rho_{1,-} & = & \lambda_{2,-} \rho_{1,+} & = & 
-\displaystyle{\frac{2\sin\theta}{2+\module{\nu}^2}}, \\
\\
\lambda_{1,+} \rho_{2,-} & = & \lambda_{1,-} \rho_{2,+} & = & 
\displaystyle{\frac{2\sin\theta}{2+\module{\nu}^2}}, \\
\\
\lambda_{3,+} \rho_{3,+} & = & \lambda_{3,-} \rho_{3,-} & = & 
-i\displaystyle{\frac{2\sin\theta\module{\nu}^2}{2+\module{\nu}^2}} .
\end{array}
\right.
\label{eq:iddirac_sta}
\end{equation}
By using the relations~\refp{eq:id_dirac_1}, \refp{eq:id_dirac_2} one can check that 
equations~\refp{eq:field_+end}, \refp{eq:field_-end} are 
the discretized version of the Dirac equation in the standard representation. 

%*****************************************************************************

\section{Conclusion and perspectives}~\label{concl}
In this article we have recovered the Dirac equation in $1+1$ space-time dimensions 
from a microscopic current model based on the Huygens principle and constructed from
suitable symmetries, thus extending to spinor waves a previous work describing the propagation
of scalar waves in an arbitrary inhomogeneous medium~\cite{detoro}. This is not surprising since 
the current model combines all the basic features of wave propagation and since the symmetries
are those underlying the Dirac equation. The originality of the method stems from 
the systematic construction based on very simple dynamical rules. Moreover the equation
derived in this paper is independent of the type of discrete lattice and could be 
derived on any graph. The method developed in this paper is very close to the philosophy 
sustaining the construction of various cellular automata which are routinely used to 
solved numerically hydrodynamic or phase transition problems. Indeed, even if we cannot 
assign a precise physical meaning to the current variables, 
we choose the dynamics of those basic variables according to the fundamental laws 
of wave propagation. Hopefuly, the current model for the 
Dirac equation could be used as a useful tool to study numerically the time-dependent 
spinor propagation in complex systems with any kind of boundary conditions. 

A natural extension of this work would be to describe spinor field propagation in 
inhomogeneous media. Indeed, a wide range of condensed matter problems are related to the 
random mass Dirac equation in $1+1$ space-time dimensions and could be studied by using such 
a model. Random mass Dirac spinors show up in the problem of a one-dimensional metal with a 
half-filled electron band and random backscattering~\cite{gogolin}, spin-ladder or 
quasi-one-dimensional spin Peierls systems~\cite{steiner}.
In order to derive a Dirac equation with a mass and a velocity varying with the space coordinate,
we can use the method developed in~\cite{detoro}. The description of scalar waves in an 
inhomogeneous medium was achieved with the help of additional currents that trap a fraction 
of the ``energy'' on each node. 
Such additional currents can also be introduced for the Dirac equation.

Another extension of this work would be to construct the Dirac equation in higher space dimensions. 
This should be useful for instance to study the plateau transitions in the integer
quantum Hall effect where the Dirac equation shows up in $2+1$ space-time dimensions~\cite{ho}.
Though the calculations are rather cumbersome, the Dirac equation in $3+1$ space-time dimensions
can also be constructed with our method.

Finally, it would be interesting to see whether the current model described in this paper 
could be useful as a starting point for a path-integral formulation of the usual Dirac equation
in $1+1$ space-time dimensions~\cite{kauffman}. Despite the fact that the formulation of 
path integrals for spinor fields in terms of non-commutating variables is not new within 
the framework of quantum field theory, this description suffers from a lack 
of physical interpretation. As an example the field-theoretical description 
of a spinor field doesn't describe the way the fermion particle moves along a 
selected path as sketched by Feynman~\cite{feynman}. Some progresses in this 
direction have been done by extending the analogy between quantum mechanics 
and brownian motion to the Dirac equation~\cite{ord96,kac74,kac,jacobson}. But this point of view 
focusses on a probabilistic interpretation of the Dirac particle through
Poisson processes. On the contrary our model deals with the Dirac equation itself and
should give a direct interpretation of the paths followed by a spinor particle
between two points.   

%*****************************************************************************

\vspace{1cm}
\centerline{{\bf Acknowledgments}}

During the preparation of this work, we have benefited from discussions 
with D. Sornette. S. DTA acknowledges the hospitality of 
the Earth and Space Sciences at UCLA and wants also to thank warmly Jean-Louis 
{\sc Pichard}, at the Service de Physique de l'Etat Condens\'e du CE de Saclay, 
and Jean-Marc {\sc Luck}, at the Service de Physique Th\'eorique du CE de Saclay, 
for hospitality when the paper has been completed. We are also indebted to Jean-Marc 
{\sc Luck} for a careful reading of the manuscript.  

%*****************************************************************************

\appendix
\renewcommand{\theequation}{A\arabic{equation}}
\section*{Appendix:~Construction of the Schr\"odinger equation with a spin}
\setcounter{equation}{0}
\label{app}

This appendix deals with the construction of a Schr\"odinger equation for a quantum
particle with a spin $1/2$ in $1+1$ space-time dimensions. This equation reads

\begin{equation}
i\hbar\frac{\partial\Psi}{\partial t} = \left[ 
\left(-\frac{\hbar^2}{2m}\frac{\partial^2}{\partial x^2}+v'\right)\mbox{Id} + v\sigma_1\right]\Psi,
\label{eq:cont_schrodspin}
\end{equation}
where $\Psi=(\psi_L \:\: \psi_R)^{T}$ is a two-component field. In 
section~\ref{scatt_maj}, we have obtained two discretized propagation equations for 
the chiral fields (Eqs.\ \refp{eq:field_Lmod4}, \refp{eq:field_Rmod4}) where the order 
of the discrete spatial derivative is controlled by the ratio 
$(\lambda_{2,L}/\lambda_{1,L})^{\ast}/(\lambda_{2,L}/\lambda_{1,L})$. 
To derive the Dirac equation, we have chosen $\lambda_{2,L}/\lambda_{1,L}$ to be 
imaginary in order to obtain a first order spatial derivative. If we assign 
$\lambda_{2,L}/\lambda_{1,L}$ to be real then a second order spatial derivative arises.
We write $\lambda_{2,L}/\lambda_{1,L}=\varepsilon_{\lambda}$ where $\varepsilon_{\lambda}=\pm 1$. 
The propagation equations~\refp{eq:field_Lmod4}, \refp{eq:field_Rmod4} depend
now on the two signs $\varepsilon\equiv\varepsilon_{\lambda}\varepsilon_{\mu}$, 
$\varepsilon_{\gamma}$, and on the two real parameters $\module{\nu}$, $\theta$.
Then, we recognize the discretized version of the Schr\"odinger equation with 
a spin~\refp{eq:cont_schrodspin} if we fix $\varepsilon=-1$ in the two wave propagation equations

\begin{eqnarray}
\lefteqn{ \frac{i}{2\tau}\left(\psi_L(n,t+\tau) - \psi_{L}(n,t-\tau)\right) =} \nonumber \\ 
&&  	 -\left(\frac{\sin\theta}{2+{\module{\nu}}^{2}}\right)
        \frac{a^2}{\tau}\left(\frac{\psi_{L}(n+a,t)+\psi_{L}(n-a,t)
	-2\psi_{L}(n,t)}{a^2}\right)  \nonumber \\
&&	+\frac{\varepsilon_{\gamma}}{\tau}\left(
	\frac{\sin\theta\module{\nu}^{2}}{2+{\module{\nu}}^{2}}\right)
	\psi_{R}(n,t)-\frac{1}{\tau}\left(
	\frac{2\sin\theta}{2+{\module{\nu}}^{2}}\right)
	\psi_{L}(n,t),
\label{eq:schrod_L} \\
\nonumber \\
\lefteqn{ \frac{i}{2\tau}\left(\psi_R(n,t+\tau) - \psi_{R}(n,t-\tau)\right) =} \nonumber \\
&&      -\left(\frac{\sin\theta}{2+{\module{\nu}}^{2}}\right)
        \frac{a^2}{\tau}\left(\frac{\psi_{R}(n+a,t)+\psi_{R}(n-a,t)
	-2\psi_{R}(n,t)}{a^2}\right) \nonumber \\
&&	+\frac{\varepsilon_{\gamma}}{\tau}\left(
	\frac{\sin\theta\module{\nu}^{2}}{2+{\module{\nu}}^{2}}\right)
	\psi_{L}(n,t)-\frac{1}{\tau}\left(
	\frac{2\sin\theta}{2+{\module{\nu}}^{2}}\right)
	\psi_{R}(n,t),
\label{eq:schrod_R}
\end{eqnarray}
provided that

\begin{eqnarray}
\frac{\hbar}{2m} & = & \left(\frac{\sin\theta}{2+{\module{\nu}}^{2}}\right)
\frac{a^2}{\tau} = -\frac{a^2}{2} \frac{v'}{\hbar}, \label{eq:id_schrod_1} \\
\nonumber \\
\frac{v}{\hbar} &= & \left(\frac{\sin\theta\module{\nu}^2}{2+{\module{\nu}}^{2}}\right)
\frac{\varepsilon_{\gamma}}{\tau}. \label{eq:id_schrod_2} 
\end{eqnarray}
The $L,R$ scattering matrices are parametrized by three real parameters~: 
$\varepsilon_{\mu}$, $\theta$ and  $\module{\nu}$, as $\nu$ is a purely imaginary number
(using~\refp{eq:is_matrixcoef1} and \refp{eq:is_matrixcoef2}). Introducing
again $\alpha$ (Eq.\ \refp{eq:alpha}), the matrices read

\begin{equation}
\begin{array}{lclcl}
S_L & = & -\displaystyle{\frac{2\sin\theta}{2+{\module{\nu}}^{2}}}\alpha^2\left(
\begin{array}{ccc}
1 & i\alpha^2 & \nu \\
\\
i\alpha^2 & 1 & i\alpha^2\nu \\
\\
i\alpha^2\nu & \nu & i\alpha^2\nu^{2}
\end{array}
\right) & + & \mu \, \mbox{diag}\left(\varepsilon_{\mu},\varepsilon_{\mu},1\right), 
\end{array}
\label{eq:matrice_schrodL}
\end{equation}

\begin{equation}
\begin{array}{lclcl}
S_R & = & -\displaystyle{\frac{2\sin\theta}{2+{\module{\nu}}^{2}}}\alpha^2\left(
\begin{array}{ccc}
1 & i\alpha^2 & i\alpha^2\nu \\ 
\\
i\alpha^2 & 1 & \nu \\
\\
\nu & i\alpha^2\nu & i\alpha^2\nu^{2}
\end{array}
\right) & + & \mu \, \mbox{diag}\left(\varepsilon_{\mu},\varepsilon_{\mu},1\right). 
\end{array}
\label{eq:matrice_scrodR}
\end{equation}

\end{document}